\documentclass[]{AIAA}
\usepackage{amsbsy,amssymb,amsmath,amsfonts,amssymb}
\usepackage{epsfig, algorithmic, algorithm}
\usepackage[]{subfigure}
\def\beq{\begin{equation}}
\def\eeq{\end{equation}}
\def\esplit{\end{split}}
\def\beqalign{\begin{array}{rl}}
\def\eeqalign{\end{array}}
\def\Abold{\mathbf{A}}

\def\Cbold{\mathbf{C}}
\def\Dbold{\mathbf{D}}
\def\Ebold{\mathbf{E}}

\def\Gbold{\mathbf{G}}
\def\Hbold{\mathbf{H}}
\def\Ibold{\mathbf{I}}

\def\Kbold{\mathbf{K}}

\def\Mbold{\mathbf{M}}

\def\Pbold{\mathbf{P}}

\def\Sbold{\mathbf{S}}

\def\Zbold{\mathbf{Z}}

\def\ebold{\mathbf{e}}
\def\fbold{\mathbf{f}}

\def\ubold{\mathbf{u}}
\def\vbold{\mathbf{v}}
\def\wbold{\mathbf{w}}
\def\xbold{\mathbf{x}}

\def\zbold{\mathbf{z}}

\def\etabold{\boldsymbol{\eta}}

\def\sigmabold{\boldsymbol{\sigma}}

\def\mubold{\boldsymbol{\mu}}
\def\phibold{\boldsymbol{\phi}}

\def\Omegabold{\mathbf{\Omega}}

\def\0bold{\boldsymbol{0}}
\def\0{\mathbf{\0}}

\begin{document}

\title{Linearized Aeroelastic Computations in the Frequency Domain Based on Computational Fluid Dynamics}
\author{David Amsallem\footnote{Department of Aeronautics and Astronautics}, Daniel Neumann\footnote{Department of Aeronautics and Astronautics}, Youngsoo Choi\footnote{Department of Aeronautics and Astronautics}, Charbel Farhat\footnote{Department of Aeronautics and Astronautics, Department of Mechanical Engineering, Institute for Computational and Mathematical Engineering, AIAA Fellow.}}
\affiliation{Stanford University, Stanford, CA 94305-4035, USA}

\begin{abstract}
An iterative, CFD-based approach for aeroelastic computations in the frequency domain is presented. The method relies on a linearized formulation of the aeroelastic problem and a fixed-point iteration approach and enables the computation of the eigenproperties of each of the wet aeroelastic eigenmodes. Numerical experiments on the aeroelastic analysis and design optimization of two wing configurations illustrate the capability of the method for the fast and accurate aeroelastic analysis of aircraft configurations and its advantage over classical time-domain approaches. 
\end{abstract}

\maketitle

\section{Introduction}
Computational aeroelasticity~\cite{dowell14} has traditionally relied on linear methods such as the doublet lattice method~\cite{hassig71} and the piston theory~\cite{ashley56}. While these two approaches enable accurate aeroelastic predictions in the subsonic and supersonic regimes, respectively, they fail to produce an accurate description of flutter in the transonic domain. Approaches based on computational fluid dynamics (CFD) and Arbitrary-Lagrangian-Eulerian (ALE)  formulations, have subsequently been able to produce very accurate flutter predictions in the transonic domain~\cite{geuzaine03}. 

When used in the time-domain, CFD-based aeroelastic predictions typically rely on a post processing step~\cite{juang85} that extracts the aeroelastic characteristics of the wet structure, such as damping ratio and frequencies. This post processing step can however be inaccurate and miss aeroelastic characteristics associated with certain modes altogether. Furthermore, depending on the initial conditions of the time simulation, certain modes may not be excited.

In the present work, an alternative approach based on linearized CFD and defined in the frequency domain is developed. It shares some similarities with the fixed-point iteration approach used in the doublet lattice method that can extract flutter frequencies. However, because it relies on CFD, it is accurate in the transonic regime and can also extract damping ratios for non-flutter points. As such, it provides the capability of leading to accurate aeroelastic predictions at any operating point. In turn, it will be demonstrated in this paper that such a procedure can be easily integrated in a design optimization procedure under aeroelastic constraints. 

This paper is organized as follows. Section~\ref{sec:CFD} presents the framework of CFD-based linearized aeroelasticity, focusing on the frequency domain formulation. The proposed approach for the fast computation of the system aeroelastic characteristics is developed in Section~\ref{sec:AEA}. Application to the aeroelastic analysis and design of two wing systems is presented in Section~\ref{sec:appli}. Finally, conclusions are given in Section~\ref{sec:conclu}.

\section{CFD-Based Linearized Aeroelasticity in the Frequency Domain}\label{sec:CFD}
\subsection{Three-field Formulation}
The following three-field ALE formulation is considered in this work to describe a fluid/structure interaction problem:
\begin{equation}\label{eq:PDE}
\displaystyle{\left\{ 
  \begin{array}{l l }
\displaystyle{\frac{\partial \left(\mathcal{J}\mathcal{W}\right)}{\partial t} + \mathcal{J} \nabla \cdot \left( \mathcal{F}(\mathcal{W})- \frac{\partial x}{\partial t} \mathcal{W}\right) - \mathcal{J} \nabla\cdot \mathcal{R}(\mathcal{W})}&= 0\\
\displaystyle{\rho \frac{\partial^2 \mathcal{U}}{\partial t^2} - \text{div}\left( \sigma(\epsilon(\mathcal{U}))\right) - \mathcal{B}}&= 0\\
\displaystyle{\widetilde{\rho} \frac{\partial^2 x}{\partial t^2} - \text{div}\left( \widetilde{E} : \widetilde{\epsilon}(x))\right)} &= 0.
\end{array} \right.}
\end{equation}
The first equation in~(\ref{eq:PDE}) is the ALE form of the Navier-Stokes equations, the second one, the elastodynamics equation for the structural subsystem and the third one provides the dynamics of the fluid grid, represented as a pseudo-structure. All tilde notations refer to that pseudo-structure.
In all equations, $t$ denotes time, $x(t)$ denotes the vector of time-dependent fluid grid position, $\mathcal{J}$ is the determinant of the Jacobian of $x$ with respect to the reference configuration and $\nabla$ is the gradient with respect to $x$.

$\mathcal{W}$ denotes the vector of conservative fluid variables and  $\mathcal{U}$ the structural displacements. $\mathcal{F}$ and $\mathcal{R}$ are the vectors of convective and diffusive fluxes, respectively. $\rho$ denotes material density, $\epsilon$ the strain tensor and $\sigma$ the stress tensor. $\text{div}$ is the divergence operator. $\mathcal{B}$ is a vector of volume forces acting on the structure. Finally $E$ denotes a tensor of elasticities. 

Appropriate Dirichlet and Neumann boundary conditions as well as interface conditions apply to the set of coupled partial differential equations. The interface conditions between the fluid and structural subsystems enforce the compatibility before the two velocity fields at the interface as well as the equilibrium of tractions on the wet surface. Finally continuity equations are enforced at the wet interface between the structural and fluid mesh subsystems.
The author is referred to~\cite{lesoinne01} for more details on these interface conditions.

\subsection{Semi-discretization}
The fluid subsystem is discretized in space by the finite volume method and the structural and fluid mesh subsystems discretized by the finite element method. The fluid is assumed to be inviscid in the remainder of this paper. The semi-discretization, after inclusion of the appropriate boundary conditions, leads to the following set of coupled ordinary differential equations
\begin{equation}\label{eq:ODE}
\displaystyle{\left\{ 
  \begin{array}{l l }
\dot{\widehat{\left(\Abold(\xbold)\wbold\right)}} +\phibold(\wbold,\xbold,\dot\xbold)&= \boldsymbol{0}\\
\Mbold\ddot\ubold + \fbold^{\text{int}}(\ubold,\dot\ubold)&=\fbold^{\text{ext}}(\ubold,\wbold) \\
\widetilde{\Kbold} \xbold &= \widetilde{\Kbold}_c\ubold.
\end{array} \right.}
\end{equation}
$\xbold$ is the vector of fluid mesh nodes positions and $\Abold$ is a diagonal matrix containing the corresponding cell volumes. A dot denotes a time derivative. $\wbold$ is the vector of discretized conservative fluid variables and $\ubold$ the vector of structural displacements. $\phibold$ denotes the numerical flux function. $\Mbold$ is the mass matrix resulting from the finite element discretization and $\fbold^{\text{int}}$ is the corresponding vector of internal forces.  $\fbold^{\text{ext}}$ is the vector of external forces applied to the  structure. Finally, the fluid mesh motion is modeled as being piece-wise static. $\widetilde{\Kbold}$ is the corresponding fictitious stiffness matrix and $\widetilde{\Kbold}_c$ is a transfer matrix representing the effect of the structural motion $\ubold$ at the wet interface.

\subsection{Linearization}
The set of nonlinear ODEs is then linearized around an equilibrium point $(\wbold_0,\xbold_0,\ubold_0,\dot\ubold_0)$ of the system. The resulting linearized system can then be written as a set of coupled linear ODEs
\begin{gather}\label{eq:LODE}
\begin{split}
\Abold_0 \dot\wbold + \Hbold_0 \wbold +(\Ebold_0+\Cbold_0) \dot\xbold + \Gbold_0\xbold&= \boldsymbol{0}\\
\Mbold\ddot\ubold +\Dbold_0\dot\ubold + \Kbold_0\ubold &=\Pbold_0\wbold\\
\widetilde{\Kbold} \xbold &= \widetilde{\Kbold}_c\ubold.
\end{split}
\end{gather}
where the fluid mesh variable $\xbold$ has been eliminated through $\xbold =\widetilde{\Sbold}\ubold = \widetilde{\Kbold}^{-1}
\widetilde{\Kbold}_c\ubold$ and the linearized operators are
\begin{gather}
\begin{split}
\Abold_0 &= \Abold(\xbold_0),~~\Hbold_0 = \frac{\partial \phibold}{\partial \wbold}(\wbold_0,\xbold_0,\dot\xbold_0),~~\Ebold_0 = \left[\frac{\partial \Abold}{\partial \xbold}(\xbold_0)\wbold_0\right]\widetilde{\Sbold},\\
\Cbold_0 &= \left[\frac{\partial \phibold}{\partial \dot\xbold}(\wbold_0,\xbold_0,\dot\xbold_0)\right]\widetilde{\Sbold},~~\Gbold_0 = \left[\frac{\partial \phibold}{\partial \xbold}(\wbold_0,\xbold_0,\dot\xbold_0)\right]\widetilde{\Sbold},\\
\Dbold_0 &= \frac{\partial \fbold^{\text{int}}}{\partial \dot\ubold}(\ubold_0,\dot\ubold_0),~~\Kbold_0 = \frac{\partial \fbold^{\text{int}}}{\partial \ubold}(\ubold_0,\dot\ubold_0) - \frac{\partial \fbold^{\text{ext}}}{\partial \ubold}(\wbold_0,\ubold_0),~~\Pbold_0 =\frac{\partial \fbold^{\text{ext}}}{\partial \wbold}(\wbold_0,\ubold_0).
\end{split}
\end{gather}
The variables $(\wbold,\ubold,\dot\ubold)$ denote now perturbations around the reference state $(\wbold_0,\ubold_0,\dot\ubold_0)$. For simplicity, the subscripts $_0$ are dropped in the remainder of this work.

\subsection{Structural Modal Reduction}
The dimensionality of the structural subsystem is subsequently reduced by modal truncation. For that purpose, a set of $m$ eigenpairs $\{\omega_i,\zbold_i\}_{i=1}^{m}$ satisfying the following generalized eigenvalue problem is first computed:
\begin{equation}
\Kbold \zbold = \omega^2\Mbold\zbold.
\end{equation}
The eigenpairs are gathered in a diagonal matrix $\Omegabold^2=\text{diag}(\omega_1,\cdots,\omega_{m})$ of eigenvalues and a modal basis matrix $\Zbold = [\zbold_1,\cdots,\zbold_m]$.

The structural displacements and velocities are then approximated as a linear combination of the eigenmodes as
\begin{equation}
\ubold(t) \approx \Zbold\ubold_m(t),~~\vbold(t) \approx \Zbold\vbold_m(t).
\end{equation}
The structural equations are then reduced by Galerkin projection as
\begin{equation}
\Ibold_m \ddot\ubold_m + \Omegabold^2\ubold_m = \fbold_m(t)
\end{equation}
where $ \fbold_m(t)$ is the reduced vector of linearized aerodynamic forces defined as
\begin{equation}
 \fbold_m(t) = \Zbold^T\fbold(t) = \Zbold^T\Pbold\wbold(t) = \Pbold_m\wbold(t) .
\end{equation}

\subsection{Frequency Domain Formulation}

The coupled equations resulting from the CFD-based ALE formulation can be written as
\begin{gather}
\begin{split}
\Ibold \ddot\ubold_m + \Omegabold^2\ubold_m &= \Pbold_m\wbold\\
\Abold\dot\wbold + \Hbold\wbold + (\Ebold+\Cbold)\dot\ubold_m + \Gbold\ubold_m&=\boldsymbol{0}
\end{split}
\end{gather}
The aerodynamic and aeroelastic linear  operators $\Hbold$, $\Ebold$, $\Cbold$, $\Gbold$ and $\Pbold$ are computed for specific structural configurations and free-stream boundary conditions such as free-stream Mach number $M_\infty$, angle of attack $\alpha$ as well as flight altitude $h$. 
Let $s\in\mathbb{C}$ and assume that
\begin{equation}
\wbold(t)=\wbold e^{st},~\ubold_m(t) = \ubold_m e^{st}.
\end{equation}
Note that $s$ can have a complex part. Then their derivatives are
\begin{equation}
\dot\wbold(t)=s\wbold e^{st},~\dot\ubold_m(t) = s\ubold_m e^{st},
\end{equation}
and the fluid equation becomes
\begin{equation}
\left(s\Abold+\Hbold\right) \wbold + \left[ s(\Ebold+\Cbold)+\Gbold\right]\ubold_m = \boldsymbol{0}.
\end{equation}
Hence
\begin{equation}
\wbold = - \left(s\Abold+\Hbold\right)^{-1} \left[ s(\Ebold+\Cbold)+\Gbold\right]\ubold_m.
\end{equation}
Plugging this expression for $\wbold$ into the structural equation leads to
\begin{equation}
\Ibold\ddot\ubold_m + \Omegabold^2\ubold_m = -  \Pbold  \left(s\Abold+\Hbold\right)^{-1} \left[ s(\Ebold+\Cbold)+\Gbold\right]\ubold_m
\end{equation}
Defining the matrix of generalized aerodynamic forces $\mathcal{A}(s) =   \Pbold  \left(s\Abold+\Hbold\right)^{-1} \left[ s(\Ebold+\Cbold)+\Gbold\right]$, this leads to
\begin{equation}\label{eq:ODE2}
\boxed{ \Ibold\ddot\ubold_m + \left(\Omegabold^2+\mathcal{A}(s)\right)\ubold_m = \boldsymbol{0}.}
\end{equation}
Note that this equation is also valid for  values of $s$ with non-zero real part.


\section{Aeroelastic Eigen Analysis}\label{sec:AEA}
\subsection{Nonlinear Aeroelastic Eigenproblem}

Consider now a motion 
\begin{equation}
\ubold_m(t) = \ubold_h e^{st}
\end{equation}
where $s = \overline\omega(i+\gamma) = i\overline\omega + \overline\omega\gamma$. Then, $\ddot\ubold_m = s^2\ubold_h e^{st} = s^2\ubold_m$ and (\ref{eq:ODE2}) becomes
\begin{equation}\label{eq:uhEq}
\boxed{\left[ s^2\Ibold + \Omegabold^2 + \mathcal{A}(s)\right] \ubold_h = \boldsymbol{0}.}
\end{equation}

This equation will now be transformed into state-space form. For that purpose, consider the velocity $\vbold_m(t) = \vbold_he^{st}$. Then $\vbold_h = s\ubold_h$. 
The operator $\mathcal{A}(s)$ can also be decomposed into its real and imaginary parts as
\begin{equation}
\mathcal{A}(s) = \mathcal{A}^R(s) + i \mathcal{A}^I(s). 
\end{equation}
Let $s$ be also decomposed into its real and imaginary parts $s=s^R+is^I$ (note that $s^R=\overline\omega\gamma$ and $s^I=\overline\omega$). Then
\begin{gather}
\begin{split}
\mathcal{A}(s) &= \mathcal{A}^R(s) + i \mathcal{A}^I(s) \\
&=  \mathcal{A}^R(s) - s^R \frac{\mathcal{A}^I(s)}{s^I} + s^R \frac{\mathcal{A}^I(s)}{s^I}   + i s^I \frac{\mathcal{A}^I(s)}{s^I} \\
&=\left( \mathcal{A}^R(s) - s^R \frac{\mathcal{A}^I(s)}{s^I}  \right) + s  \frac{\mathcal{A}^I(s)}{s^I}.
\end{split}
\end{gather}
Equation~(\ref{eq:uhEq}) can then be written as
\begin{equation}
s\Ibold(s\ubold_h) + \Omegabold^2\ubold_h + \left( \mathcal{A}^R(s) - s^R \frac{\mathcal{A}^I(s)}{s^I}  \right)\ubold_h +   \frac{\mathcal{A}^I(s)}{s^I} (s\ubold_h) = \boldsymbol{0}
\end{equation}
which, combined with $\vbold_h=s\ubold_h$ leads to the system
\begin{equation}
\displaystyle{\left\{ 
  \begin{array}{l l }
s\Ibold\vbold_h + \left[ \Omegabold^2 + \mathcal{A}^R(s) - \frac{s^R}{s^I}\mathcal{A}^I(s)\right]\ubold_h + \frac{\mathcal{A}^I(s)}{s^I}\vbold_h &= \boldsymbol{0}, \\
\vbold_h &=s\ubold_h
\end{array} \right.}
\end{equation}
that can be written in matrix form as
\begin{equation}
\left[\begin{array}{c|c}\boldsymbol{0} & \Ibold \\\hline -\left(\displaystyle{\Omegabold^2 + \mathcal{A}^R(s) - \frac{s^R}{s^I}\mathcal{A}^I(s)}\right) & \displaystyle{ - \frac{\mathcal{A}^I(s)}{s^I}}\end{array}\right]\left[\begin{array}{c}\ubold_h \\\hline \vbold_h\end{array}\right]=s\left[\begin{array}{c}\ubold_h \\\hline \vbold_h\end{array}\right].
\end{equation}
This leads to the following non-linear eigenvalue problem
\begin{equation}\label{eq:NLEig}
\mathbb{A}(s) \left[\begin{array}{c}\ubold_h \\\hline \vbold_h\end{array}\right]=s\left[\begin{array}{c}\ubold_h \\\hline \vbold_h\end{array}\right]
\end{equation}
where $\mathbb{A}(s)$ is the real-valued matrix
\begin{equation}
\mathbb{A}(s) =\left[\begin{array}{c|c}\boldsymbol{0} & \Ibold \\\hline -\left(\displaystyle{\Omegabold^2 + \mathcal{A}^R(s) - \frac{s^R}{s^I}\mathcal{A}^I(s)}\right) & \displaystyle{ - \frac{\mathcal{A}^I(s)}{s^I}}\end{array}\right].
\end{equation}
Note that there are as many solutions to this eigenvalue problem as there are generalized degrees of freedom in the structural system. The eigenvalues come in pairs $s_j  =s_j^R \pm i s_j^I,~j=1,\cdots,m$.

There are two main approaches that can be followed to solve the non-linear eigenvalue problem~(\ref{eq:NLEig}).

The first class of approaches is based on a fixed point iteration approach. In the context of the p-k method for solving the flutter problem, that approach was first introduced in~\cite{hassig71}. At each iteration, a linear eigenvalue system is solved and the algorithm proceeds until convergence of the iterates to an eigenvalue. The advantage of that approach is that it relies only on a simple linear eigenvalue solver and is therefore straightforward to implement. A drawback of this approach is the potential slow convergence requiring many iterations~\cite{back97}. ~\cite{back97} presents an analysis of the convergence  proposes an extension based on Newton's method to address the slow convergence issue. That extension however requires the computational of the sensitivities of the aeroelastic operators with respect to the reduced frequency.

The second class of approaches is based on continuation to solve the non-linear eigenvalue problem. Such an approach was proposed in~\cite{meyer88} in the context of the p-k method. The main advantage of this method is its robustness that guarantees a computation of  all the eigenvalues. However, this approach is expensive as it requires computing  flutter predictions for an entire trajectory from high altitude to the desired altitude. Further, computing the flutter path in a robust fashion may require small increments that increase the computational cost. Finally, sensitivities of the aeroelastic operator are required for a continuation approach.

In the remainder of this paper, the fixed point iteration approach is chosen for its simplicity of implementation. It is described in Section~\ref{sec:FPI}.


\subsection{Fixed-point Iteration Algorithm}\label{sec:FPI}

The algorithm proceeds, for each of the $m$ structural eigenvalues, by iteratively computing one solution of the eigenvalue problem~(\ref{eq:NLEig}). For that purpose, an initial guess is provided by the dry eigenvalues and associated eigenvectors. The operator $\mathbb{A}$ is then evaluated at the dry mode and all of its eigenvalues computed. The eigenvalue are sorted according to their imaginary parts and the eigenvalue associated with the index of interest retained. The operator $\mathbb{A}$ is then evaluated for that eigenvalue. The procedure is repeated until convergence. A pseudo-code is provided  in Algorithm~\ref{alg:iterative}. If convergence is not achieved after a maximum number of allowed iterations, the continuation procedure mentioned above is triggered. 

The main difference with the p-k method developed in~\cite{hassig71} is the fact that the real part of the eigenvalues is retained during the iterations, leading to a possible convergence of the procedure to an aeroelastic eigenvalue with non-zero real part. This means that aeroelastic eigenvalue can be computed for non flutter configurations as well. This is due to the fact that the linearized CFD-based framework developed in Section~\ref{sec:CFD} is equally valid at flutter and non-flutter configurations, unlike the p-k method that is based on the doublet lattice method, valid only for harmonic motions.

\begin{algorithm}
\caption{Iterative computation of the eigenvalues of the nonlinear eigenproblem~(\ref{eq:NLEig})}
\begin{algorithmic}[1]
\REQUIRE Dry structural frequencies $\{\omega_j\}_{j=1}^m$, tolerance $\epsilon$
\ENSURE Eigenvalues $\{s_j=s_j^R+is_j^I\}_{j=1}^m,$ 
\FOR{$j=1,\cdots,m$}
\STATE Let $s_j^0 = i \omega_j$ and $k=0$
\WHILE {$k=0$ or $s_j^k-s_j^{k-1}>\epsilon$}
\STATE $k=k+1$
\STATE Compute all the eigenvalues $\{\lambda_l = \lambda_l^R\pm i \lambda_l^I\}_{j=1}^m$ of $\mathbb{A}(s_j^{k-1})$ and order them  by frequency such that
\[ \lambda_1^I < \cdots < \lambda_m^I\]
\item Let $s_j^k = \lambda_j^R + i \lambda_j^I$
\IF{$k\geq k_{\max}$}
\STATE Call the continuation procedure 
\ENDIF
\ENDWHILE
\ENDFOR
 \end{algorithmic}\label{alg:iterative}
\end{algorithm}
Once the eigenvalues $\{s_j=s_j^R+is_j^I\}_{j=1}^m$ of the nonlinear eigenproblem~(\ref{eq:NLEig}) are computed, the associated aeroelastic frequencies and damping ratios can be readily computed as
\begin{equation}\label{eq:damping}
\widehat{\omega}_j = s_j^I,~~\eta_j = -\frac{s_j^R}{\sqrt{(s_j^R)^2+(s_j^I)^2}},~j=1,\cdots,m.
\end{equation}

\subsection{Complexity analysis}
The main cost associated with the proposed procedure originates from the computation of the linear operator $ \mathbb{A}(s)$ for values of $s$ arising in the procedure. Computing $ \mathbb{A}(s)$ requires the computation of the matrix of generalized forces $\mathcal{A}(s) =   \Pbold  \left(s\Abold+\Hbold\right)^{-1} \left[ s(\Ebold+\Cbold)+\Gbold\right]$. In practice, computing the term $\left(s\Abold+\Hbold\right)^{-1} \left[ s(\Ebold+\Cbold)+\Gbold\right]$ is the most expensive step. It relies on the solution of $m$ large-scale, sparse systems of equations $\left(s\Abold+\Hbold\right) \xbold_i = \left[s(\Ebold+\Cbold)+\Gbold\right]\ebold_i$,  $i=1,\cdots,m$. These solutions are typically computed by a Krylov-subspace iterative technique. In the present work, the pre-conditioned Generalized Minimal Residual technique (GMRES)~\cite{saad86} is used. 

Assuming, for simplicity, that the computation of each of the $m$ aeroelastic eigenvalues requires $N_\text{iter}$ iterations, $N_\text{iter} m^2$ such solutions of linear systems are required.

\subsection{Computation of the Flutter Speed Index}
The Flutter Speed Index (FSI) is a non-dimensional quantity that characterizes the flutter properties of an aeroelastic system. It is computed for a fixed free-stream Mach number $M_\infty$ and density $\rho_\infty$. The free-stream pressure $p_\infty$ is increased until the onset of flutter. This is equivalent, for a perfect gas, to increasing the free-stream velocity $V_\infty$ as
\begin{equation}
V_\infty = \sqrt{\frac{\gamma p_\infty}{\rho_\infty}}M_\infty
\end{equation}
with $\gamma=1.4$. The FSI is then defined for the critical value of flutter velocity $V_\infty^{\text{cr}}$ as
\begin{equation}
\text{FSI} = \frac{V_\infty^{\text{cr}}}{b_S \omega_\alpha \sqrt{\mu}}
\end{equation}
where $b_S$ is a reference length such as the semi-chord of the wing at the root, $\omega_\alpha$ is the frequency of the first torsional dry mode of the structural system. The parameter $\mu$ is defined as
\begin{equation}
\mu = \frac{m_S}{\rho_\infty \widehat{V}}
\end{equation} 
where $m_S$ is the structural mass of the system and $\widehat{V}$ is the volume of a conical frustum with lower and upper based diameters being the stream-wise chords at the root and the tip of the wing and with height equal to the semi-span of the wing of the system.

Flutter can be in practice detected from the outputs of Algorithm~\ref{alg:iterative} by the presence of an eigenvalue $s_j$ of the operator $\mathbb{A}$ with positive real part. In practice, Algorithm~\ref{alg:iterative} is run for increasing values of the free-stream pressure until flutter is detected.

\section{Applications}\label{sec:appli}
\subsection{Flutter Analysis of the Agard Wing 445.6}
\subsubsection{Agard Wing 445.6 model}
\begin{figure}[h!]
\centering
\includegraphics[width=0.8\textwidth]{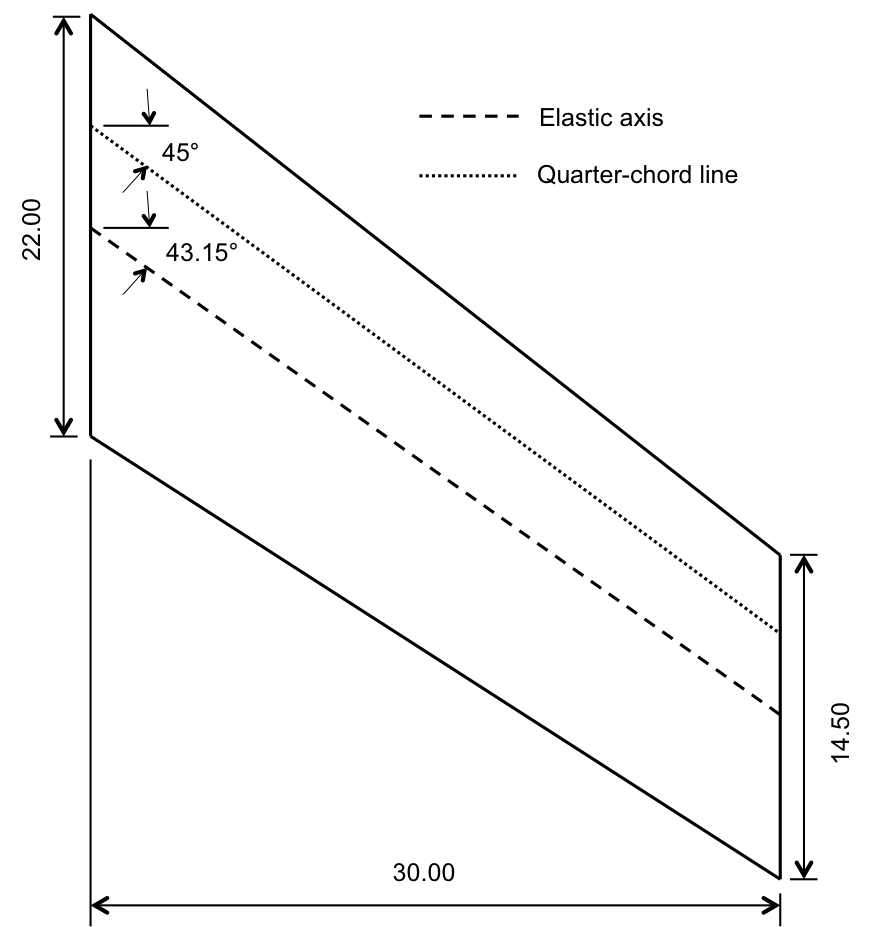}
\vspace{3mm}
\caption[]{
           Geometry of the Agard wing           
}\label{fig:agardWing} 
\end{figure}

The weakened Agard 445.6 model 3 research wing model is described in~\cite{yates87}. The dimensions of the wing are reported in Figure~\ref{fig:agardWing}. A finite element model (FEM) consisting of $800$ shell elements, corresponding to $2646$ degrees of freedom (dofs) is created and the eigenfrequencies associated with its first $m=4$ structural modes compared in Table~\ref{tab:agardFreq} to the ones measured experimentally in~\cite{yates87}. Good agreements are observed between the two sets of eigenfrequencies. 

A finite volume mesh with more than $22,000$ nodes is then created, leading to more than $110,000$ dofs in the CFD model.

\begin{table}[htdp]
\begin{center}
\begin{tabular}{lclclclc|c|}
\hline
\hline
Mode  & Frequency (Hz) & Frequency (Hz) & Relative  \\
number & (experimental model in~\cite{yates87}) & (present FEM model) & discrepancy \\
\hline
$1$&$9.60$ & $9.84$  & $2.5\%$\\
$2$& $38.17$& $39.54$ & $3.6\%$\\
$3$&$ 48.35$& $50.50$ & $4.4\%$\\
$4$& $91.54$& $96.95$ & $5.9\%$\\
\hline
\hline
\end{tabular}
\end{center}
\caption{AGARD Wing 445.6  retained modes for the structure}\label{tab:agardFreq}
\end{table}

\subsubsection{Flutter Analysis}
\begin{figure}[h!]
\vspace{-3mm}
\centering
\subfigure[$M_{\infty}=0.8$]{
\includegraphics[width=0.45\textwidth]{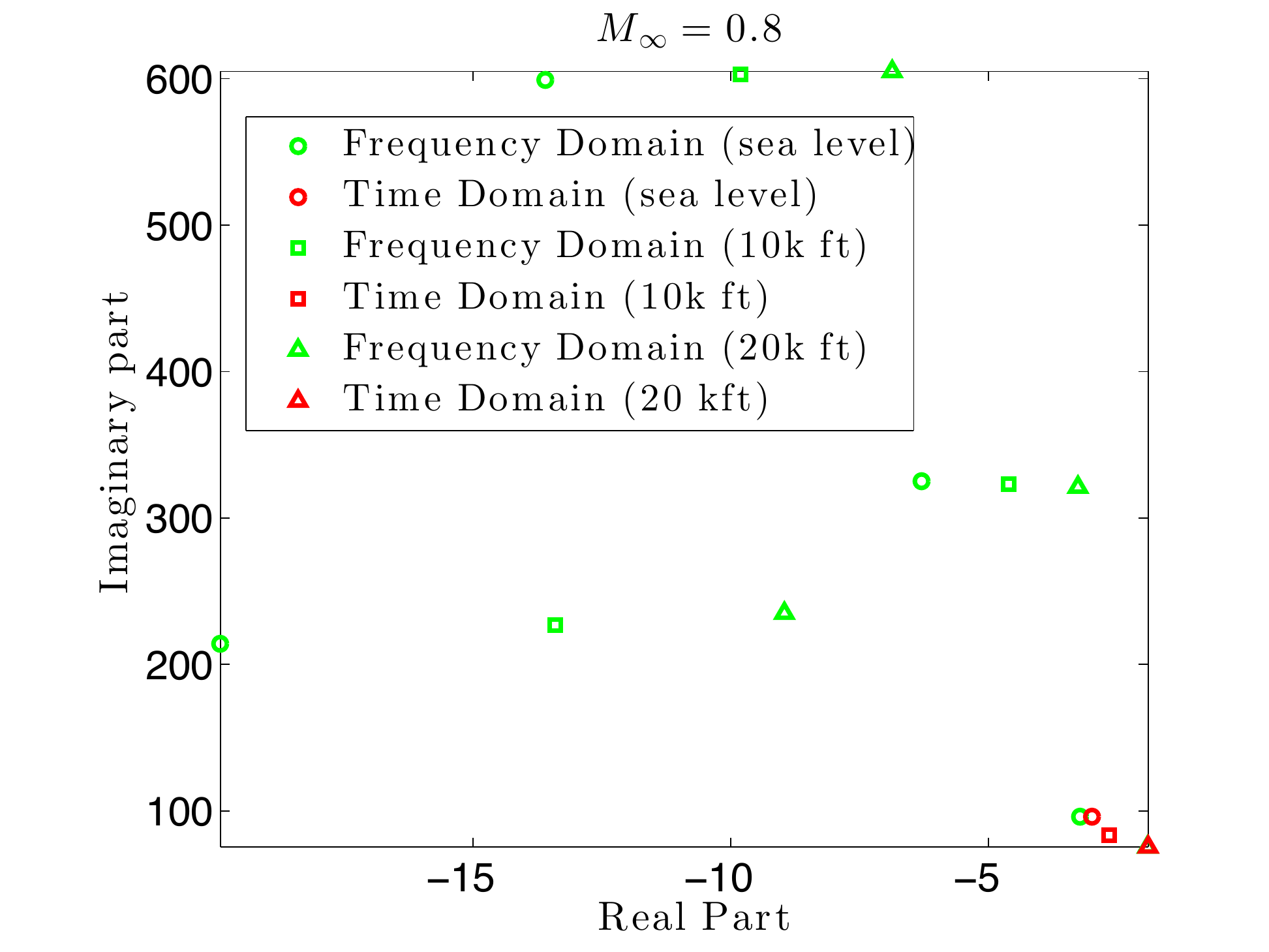}
}
\subfigure[$M_{\infty}=0.95$]{
\includegraphics[width=0.45\textwidth]{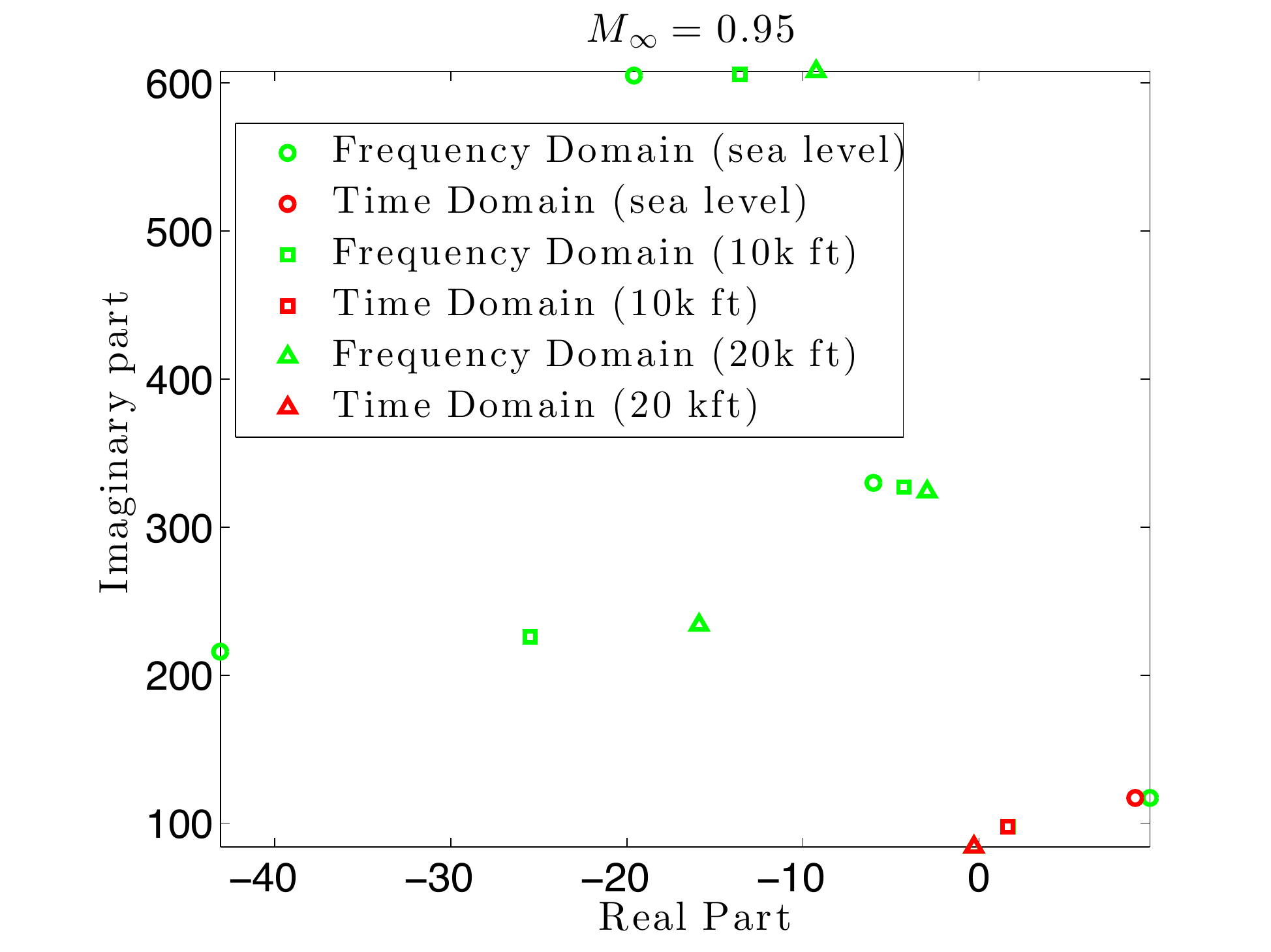}
}
\subfigure[$M_{\infty}=1.1$]{
\includegraphics[width=0.45\textwidth]{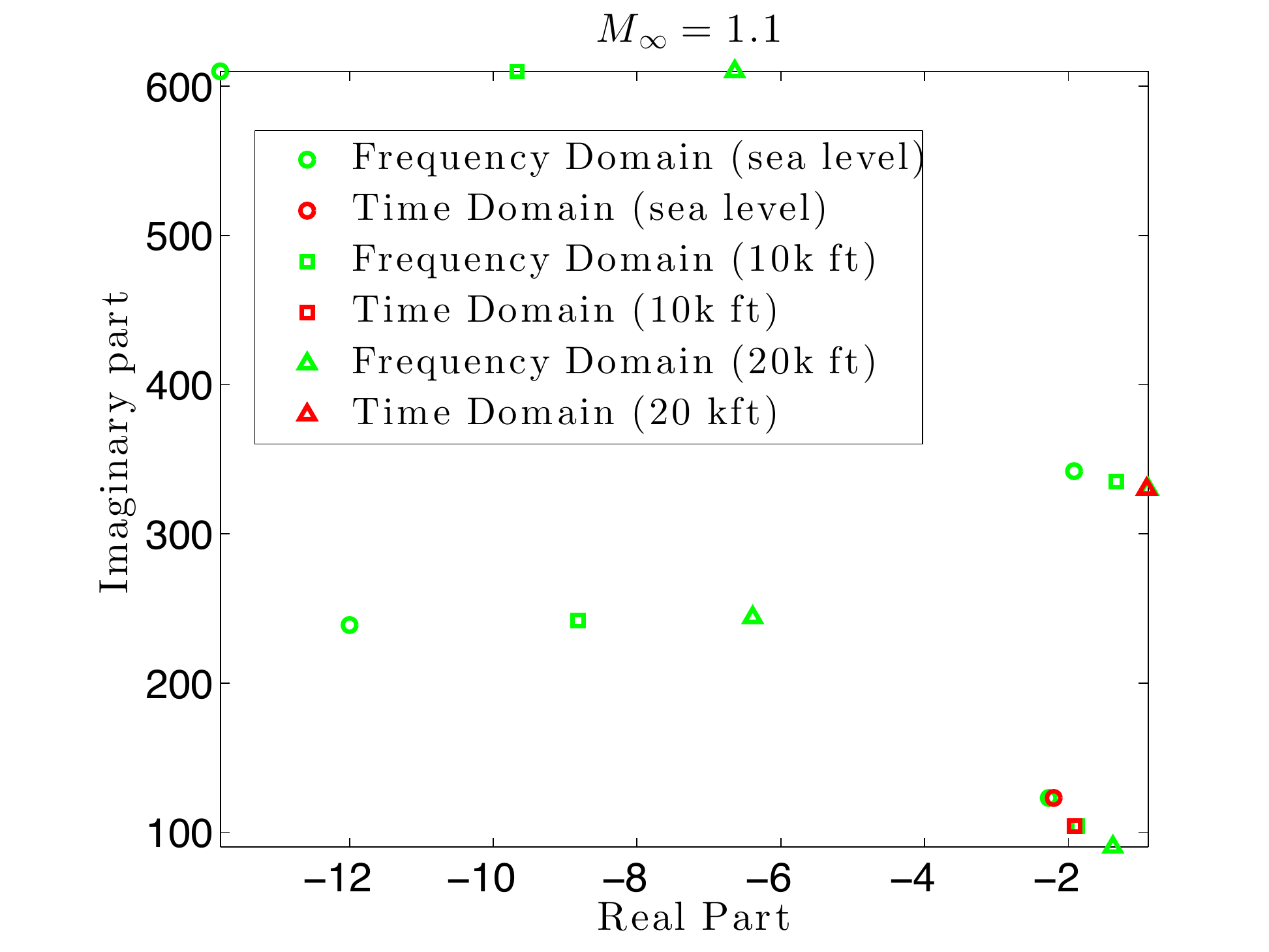}
}
\vspace{3mm}
\caption[]{
           Comparison of the aeroelastic eigenvalues predicted by the frequency and time domain methods for the Agard wing
}\label{fig:agardEvs}
\end{figure}

\begin{figure}[h!]
\vspace{-3mm}
\centering
\subfigure[$M_{\infty}=0.8$]{
\includegraphics[width=0.45\textwidth]{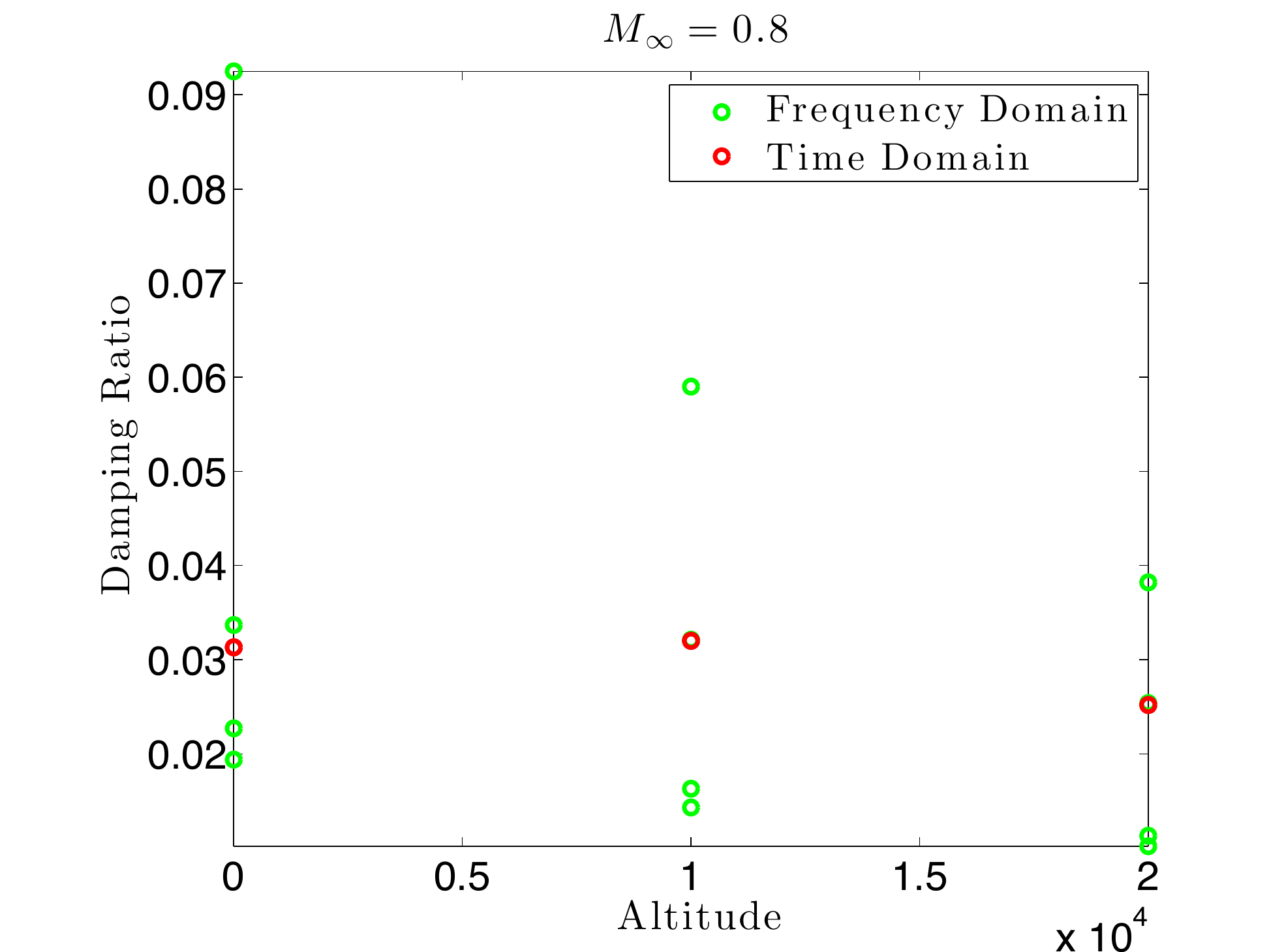}
}
\subfigure[$M_{\infty}=0.95$]{
\includegraphics[width=0.45\textwidth]{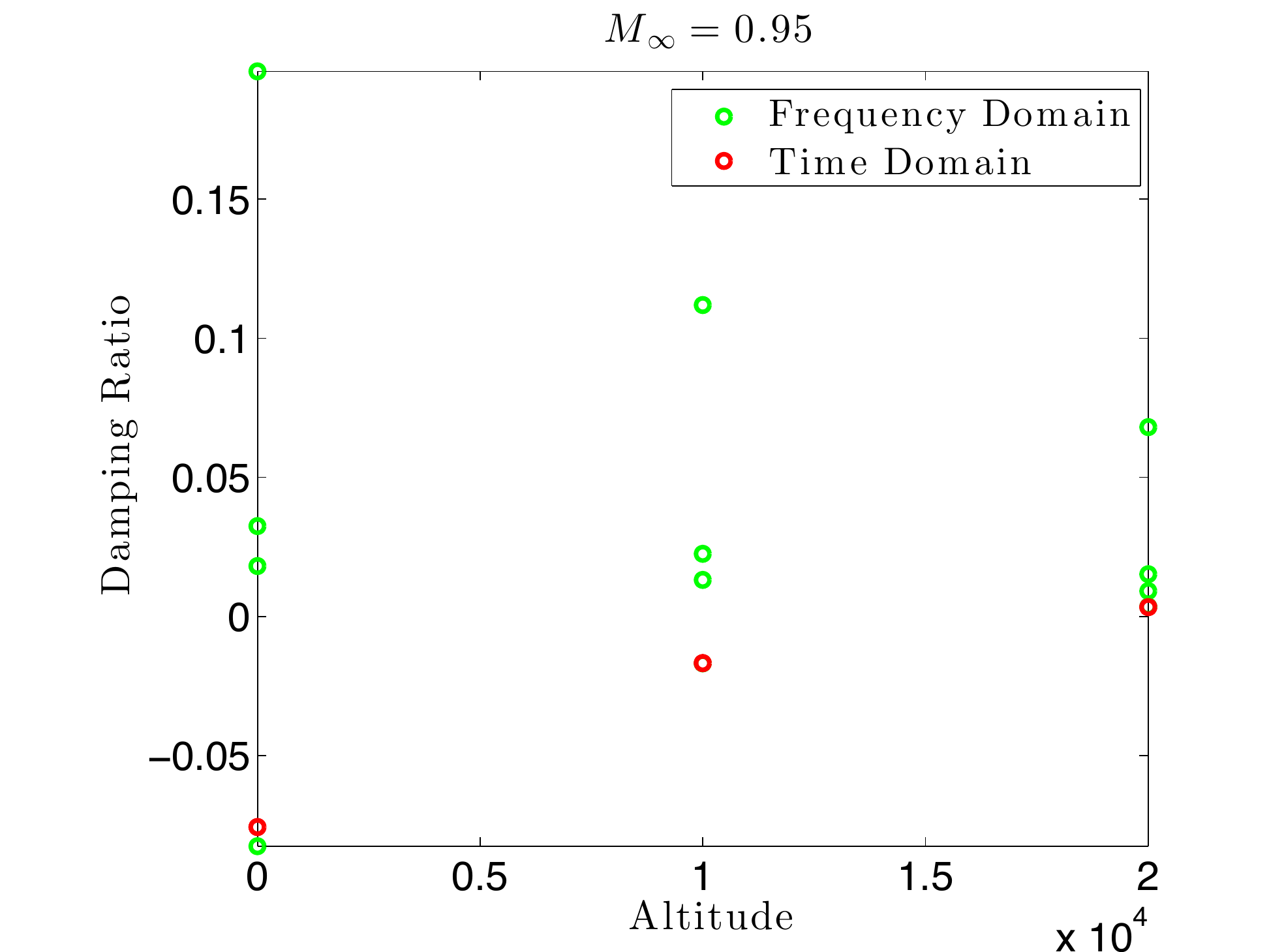}
}
\subfigure[$M_{\infty}=1.1$]{
\includegraphics[width=0.45\textwidth]{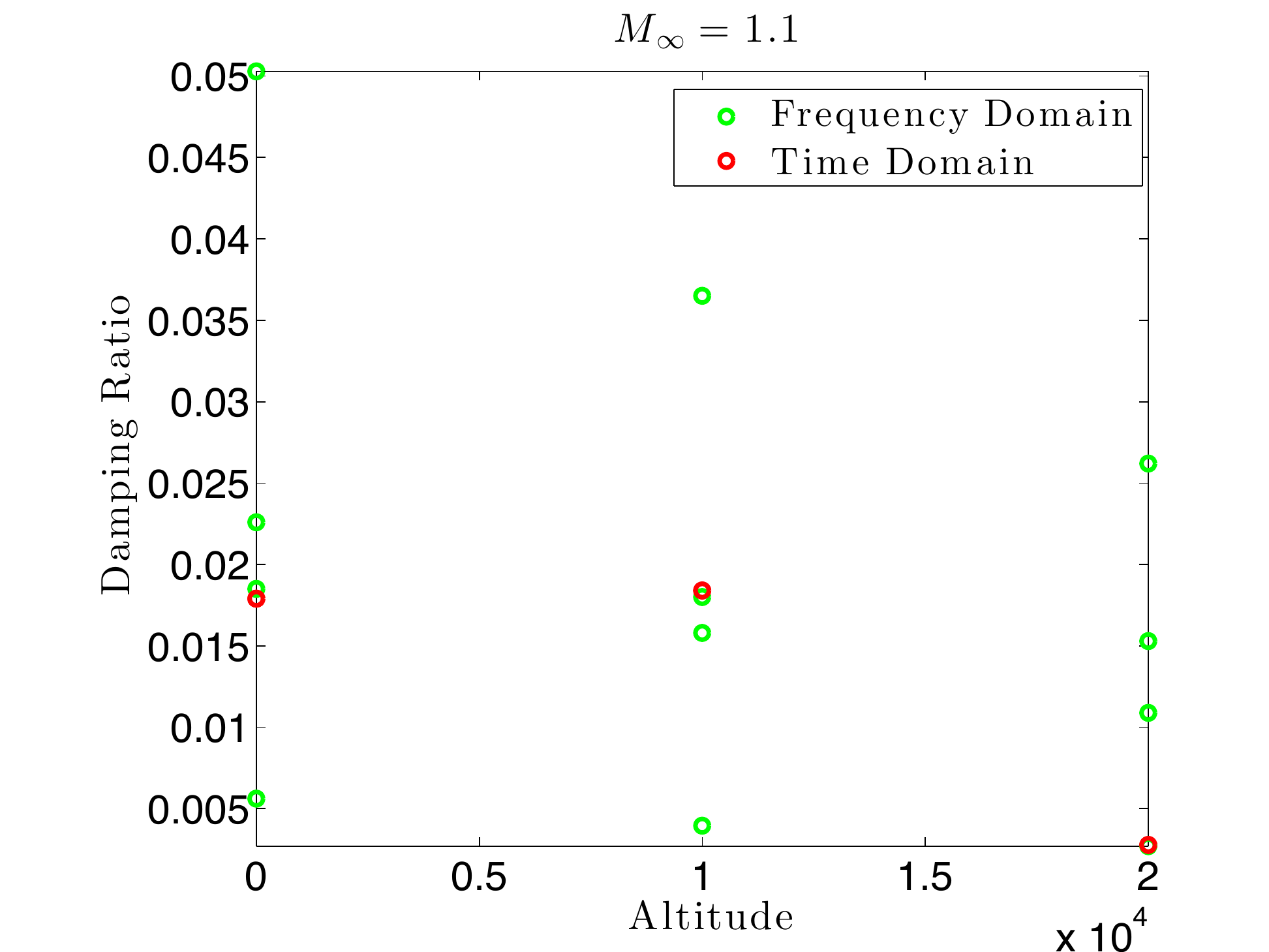}
}
\vspace{3mm}
\caption[]{
           Comparison of the smallest aeroelastic damping ratios predicted by the frequency and time domain methods for the Agard wing
}\label{fig:agardDamp} 
\end{figure}

A flutter analysis is first carried on for three different Mach numbers $M_\infty\in\{0.8,0.95,1.1\}$ and three altitudes $h\in\{0,10000,20000\}$ ft. For every combination of flight conditions, the proposed iterative approach is applied to predict the aeroelastic eigen-characteristics of the system. Hence, for each case, four sets of real and imaginary parts for the eigenvalues are computed, as well as the corresponding damping ratio, as defined in Eq.~(\ref{eq:damping}). The densities and pressures associated with each altitude are given in Table~\ref{tab:atmosphere}. The results are graphically depicted in Figure~\ref{fig:agardEvs}(a) for aeroelastic predictions at sea level, in Figure~\ref{fig:agardEvs}(b) for $h=10,000$ ft and in Figure~\ref{fig:agardEvs}(c) for $h=20,000$ ft. In addition, the results are compared to predictions in the time domain for which damping and frequencies characteristics are extracted by fitting a canonical response to the transient lift. This simple approach only allows the extraction of a single eigenvalue, as opposed to the approach proposed in this paper that allows the extraction of all the eigenvalues. The comparisons with the proposed iterative approach show that the time domain are able to accurately predict  the characteristics of one of the aeroelastic eigenmodes, but that eigenmode is not always associated with the smallest damping ratio. As a result, the time domain approach ofter overestimates the minimal damping ratio of the system, as observed in Figure~\ref{fig:agardDamp} where the damping ratios computed by each approach are reported. Furthermore, the time-domain mispredicts flutter for $M_\infty=0.95$ and $h=10$k ft.    On the other hand, the proposed iterative approach returns all the eigenvalues and hence the one associated with smallest damping.
The aeroelastic analysis for $M_\infty=0.95$ and $h=20,000$ ft takes 4.25 minutes on 8 processors for a tolerance $\epsilon=10^{-6}$. Computing the steady-state takes 1.1 minute and the iterative approach takes 3.15 minutes.
\begin{table}[htdp]
\begin{center}
\begin{tabular}{lclclclc|}
\hline
\hline
Altitude (ft)  & Pressure (lbf.in$^2$) & Density (lbf.s$^2$.in$^{-4}$) \\
\hline
$0$ (sea level)&$14.7$ & $1.1463\times10^{-7}$  \\
$10,000$& $10.11$ & $0.8466\times10^{-7}$  \\
$20,000$&$6.76$ & $0.6112\times10^{-7}$ \\
\hline
\hline
\end{tabular}
\end{center}
\caption{Free-stream pressures and densities for the altitudes considered}\label{tab:atmosphere}
\end{table}

\begin{figure}[h!]
\centering
\includegraphics[width=0.6\textwidth]{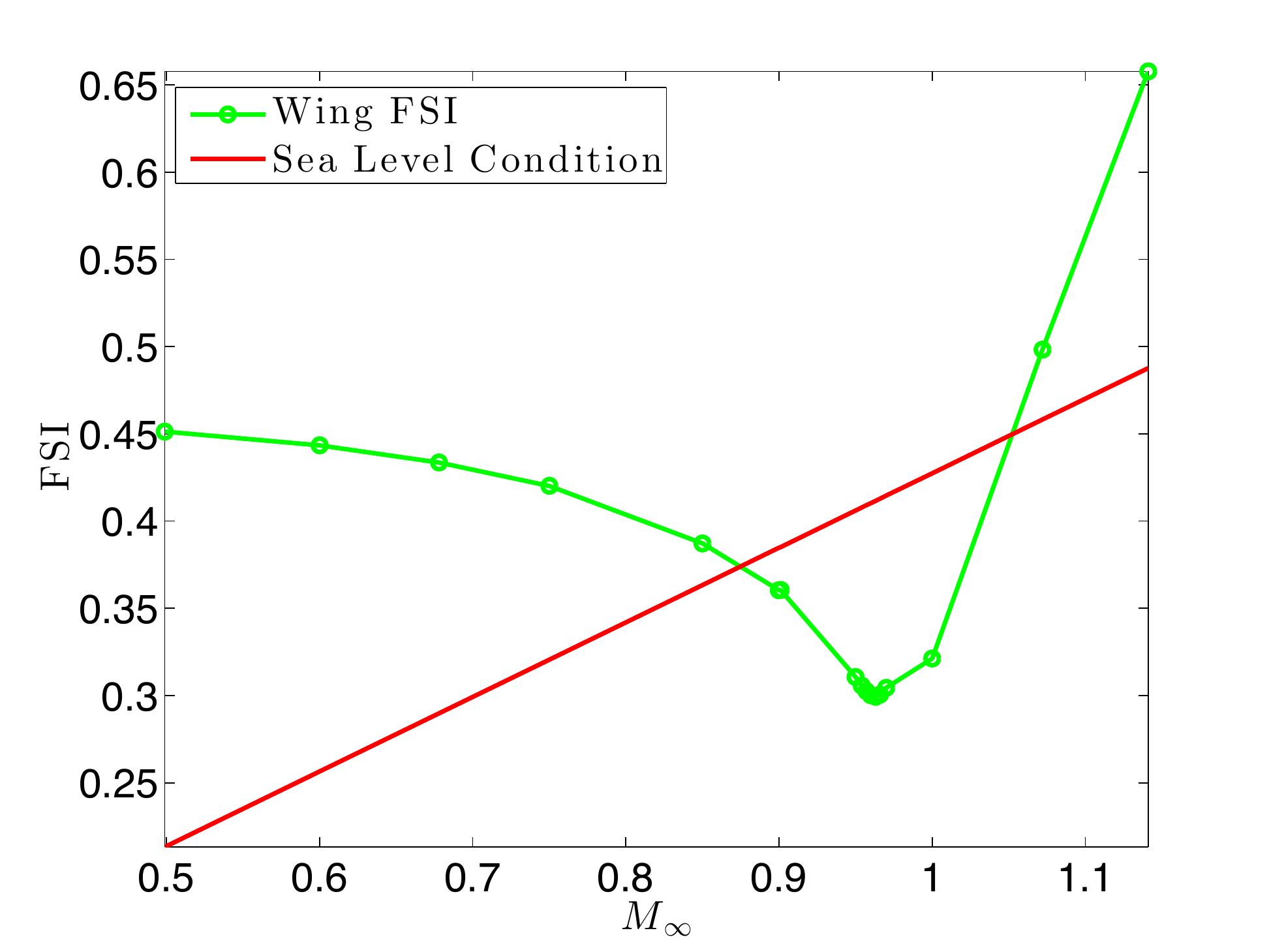}
\vspace{3mm}
\caption[]{
           Variations with the Mach number of the Flutter Speed Index and comparison with the sea level conditions for the Agard wing
}\label{fig:FSIAgard} 
\end{figure}

Next, FSIs are computed for a range of Mach numbers varying from $M_\infty=.499$ to $M_\infty=1.141$ using the proposed approach. For that purpose, the free-stream density is first fixed to its sea level value. The results are reported in Figure~\ref{fig:FSIAgard}. One can observe a characteristic flutter dip around $M_\infty= 0.96$ and a range of flutter conditions for $M_\infty\in[0.87,1.05]$.

%
%
\subsection{Analysis and Design Optimization Based on the ARW-2 Wing}
\subsubsection{Aeroelastic Research Wing ARW-2 Model}
\begin{figure}[h!]
\centering
\includegraphics[width=0.8\textwidth]{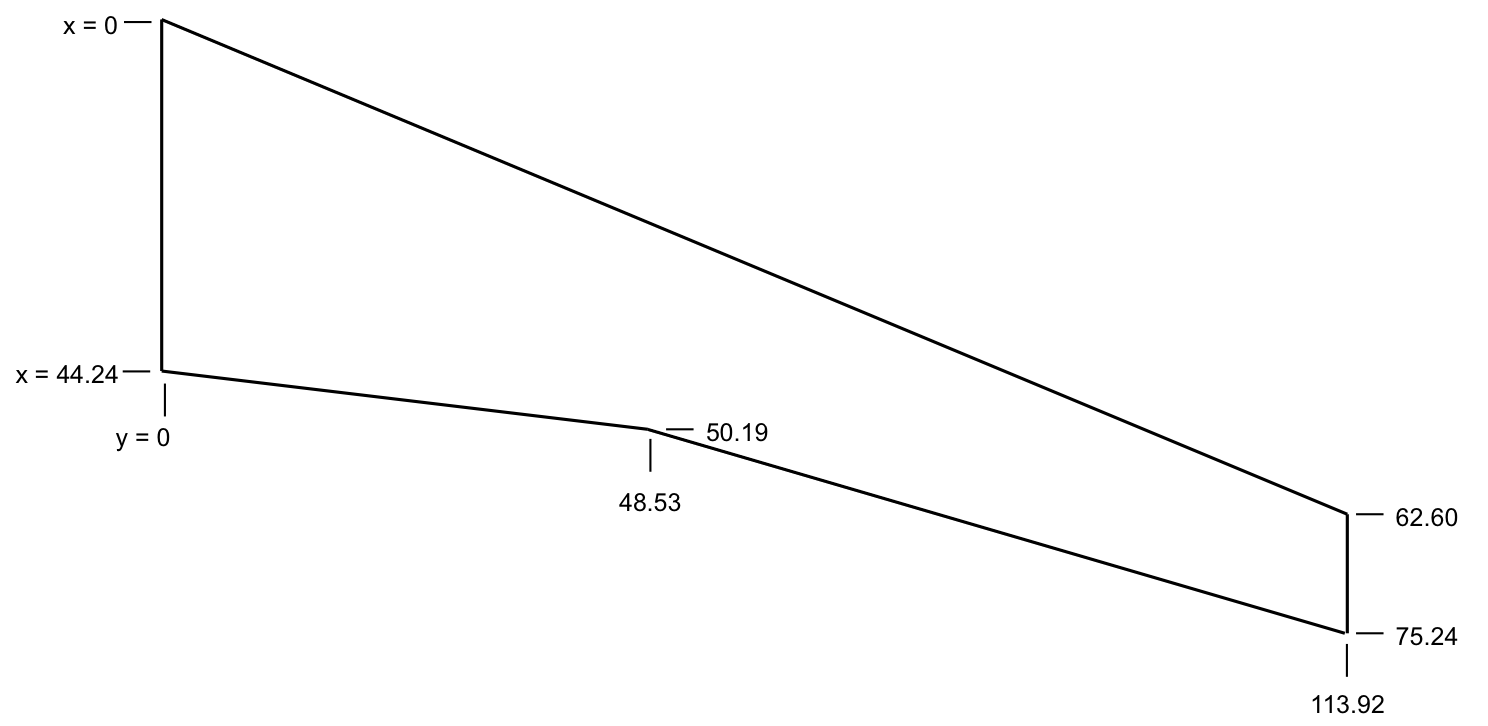}
\vspace{3mm}
\caption[]{
           Geometry of the ARW-2 wing           
}\label{fig:ARW2} 
\end{figure}

\begin{figure}[h!]
\centering
\includegraphics[width=0.8\textwidth]{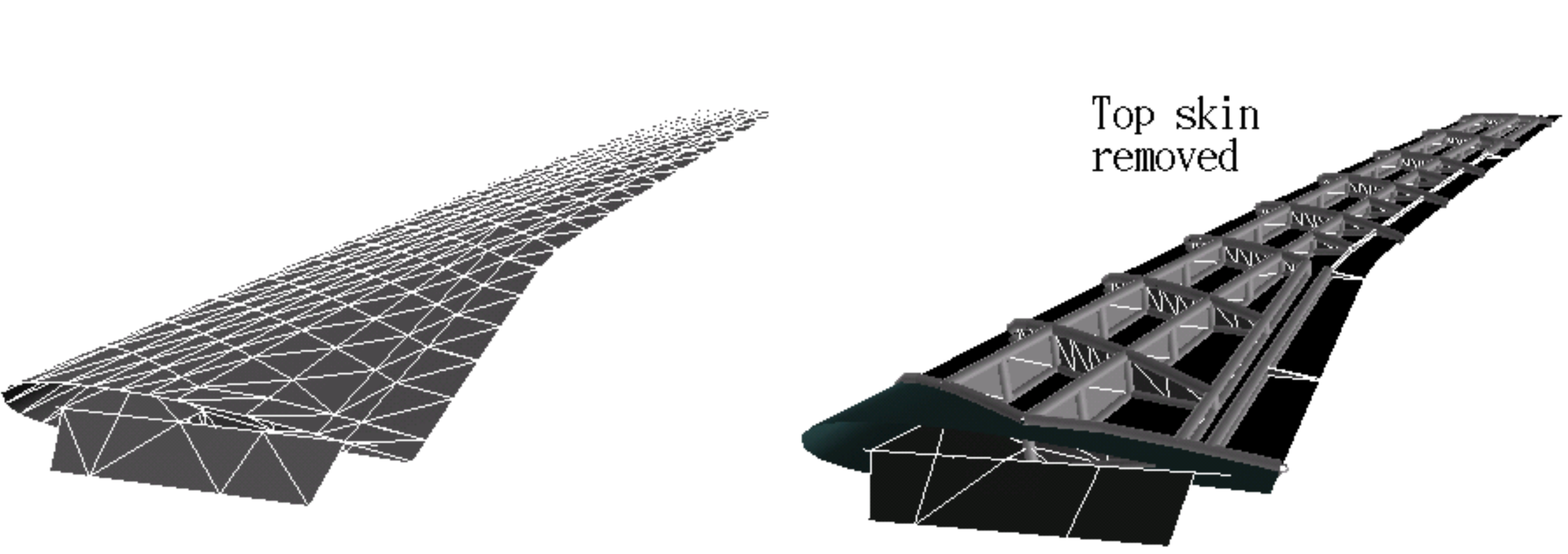}
\vspace{3mm}
\caption[]{
           Finite element model of the ARW-2 wing           
}\label{fig:ARW2_str} 
\end{figure}

The second model is a more realistic wing. It is the Aerodynamic Research Wing Number 2 (ARW-2) which was designed for flutter suppression and gust load alleviation~\cite{seidel87,sandford89}. The wing is based on a supercritical airfoil, has an aspect ratio of 10.3 and a leading-edge sweepback angle of $28.8^\circ$. The geometry of the wing is depicted in Figure~\ref{fig:ARW2}.  A FE model of the wing consists of $1765$ shell, beam and truss elements, resulting in $2736$ degrees of freedom. The structural FE model is depicted in Figure~\ref{fig:ARW2_str}. $m=8$ structural eigenmodes are then selected to model the structural subsystem. The corresponding eigenfrequencies are reported in Table~\ref{tab:arw2Freq}.

\begin{figure}[h!]
\vspace{-3mm}
\centering
\subfigure[Surface and symmetry plane]{
\includegraphics[height=0.3\textwidth]{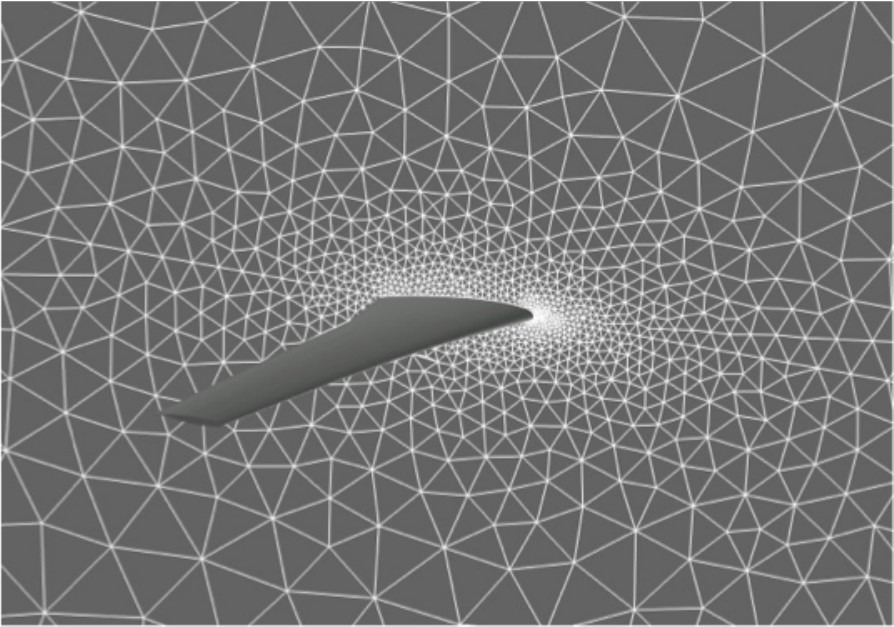}
}
\subfigure[Fluid surface mesh]{
\includegraphics[height=0.3\textwidth]{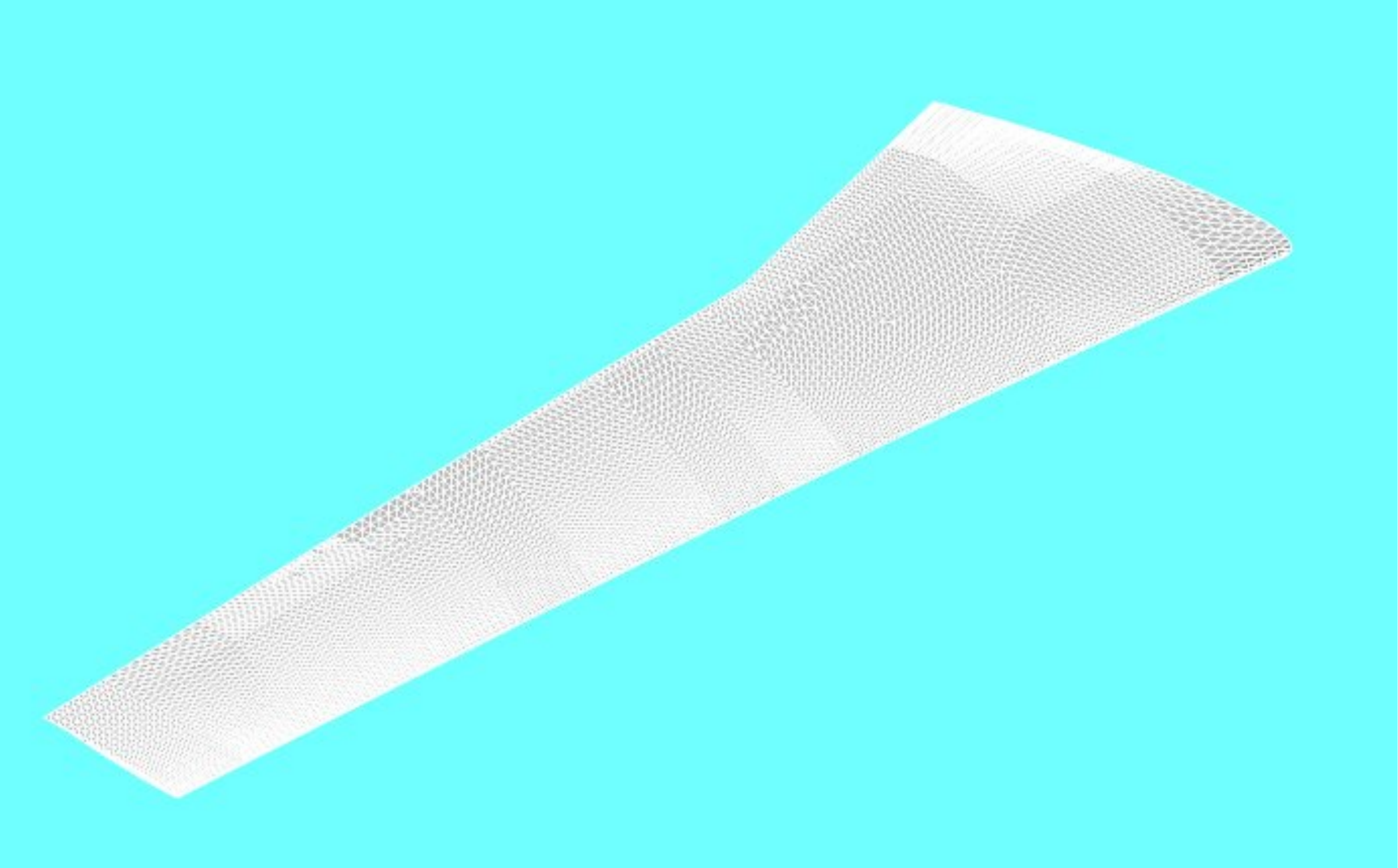}
}
\vspace{3mm}
\caption[]{
           ARW-2 wing fluid mesh
}\label{fig:ARW2_fl}
\end{figure}

The fluid computational domain consists of an unstructured grid with $47,847$ vertices corresponding to more than $260,000$ dofs. Part of the fluid mesh is shown in Figure~\ref{fig:ARW2_fl}.

\begin{table}[htdp]
\begin{center}
\begin{tabular}{lclclclc|}
\hline
\hline
Mode  number &  Frequency (Hz) \\
\hline
$1$&$7.92$   \\
$2$& $28.83$  \\
$3$&$ 33.27$ \\
$4$& $57.04$ \\
$5$&$73.70$   \\
$6$& $86.08$  \\
$7$&$ 93.91$ \\
$8$& $110.87$ \\
\hline
\hline
\end{tabular}
\end{center}
\caption{ARW2  retained modes for the structure}\label{tab:arw2Freq}
\end{table}

\subsubsection{Flutter Analysis}
A flutter analysis is carried on for the three Mach numbers $M_\infty\in\{0.8,0.95,1.1\}$ and altitudes $h\in\{0,10000,20000\}$ ft. For each combination of flight conditions, the proposed approach is applied to predict the aeroelastic eigen-characteristics of the system.

\begin{figure}[h!]
\vspace{-3mm}
\centering
\subfigure[$M_{\infty}=0.8$]{
\includegraphics[width=0.45\textwidth]{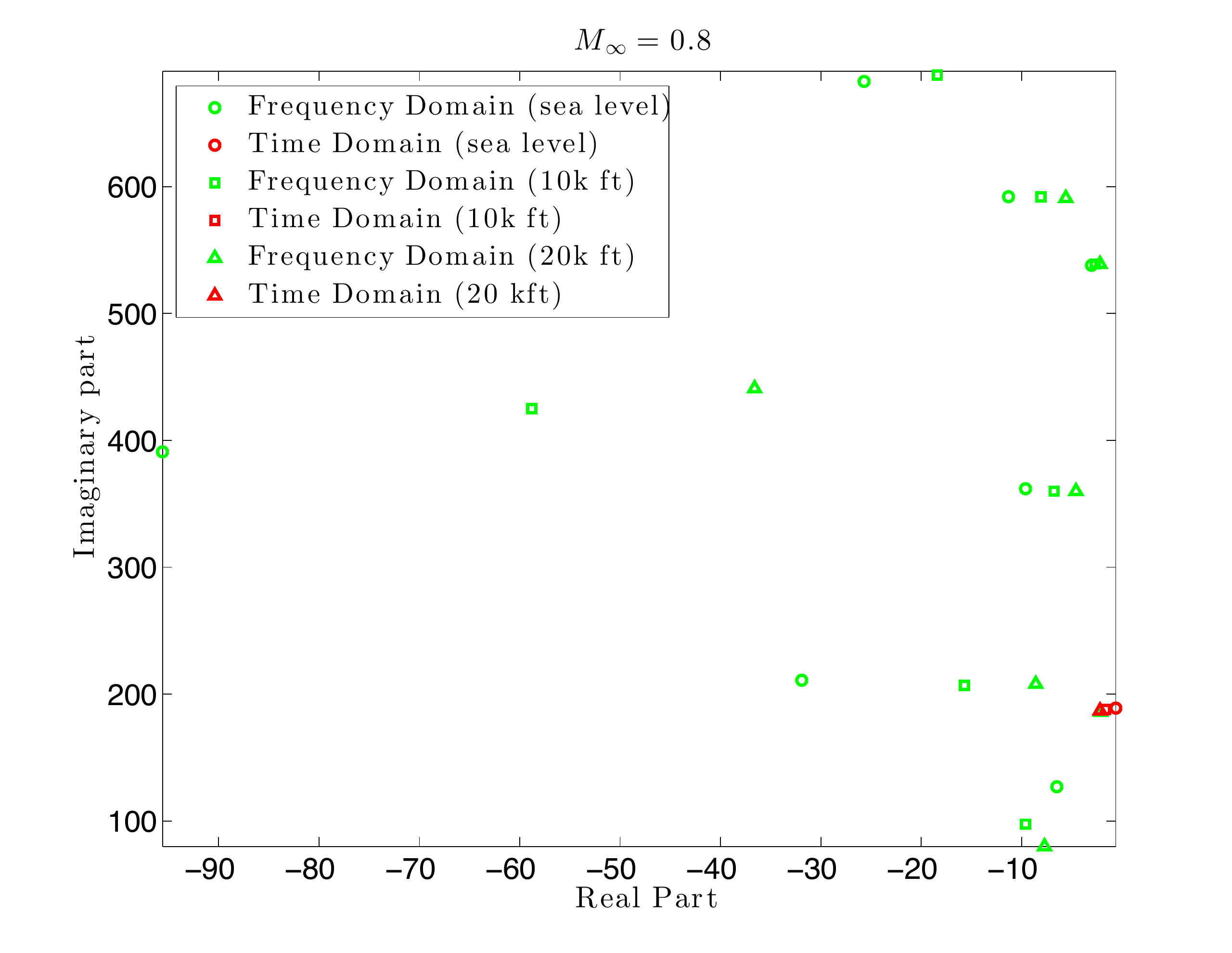}
}
\subfigure[$M_{\infty}=0.95$]{
\includegraphics[width=0.45\textwidth]{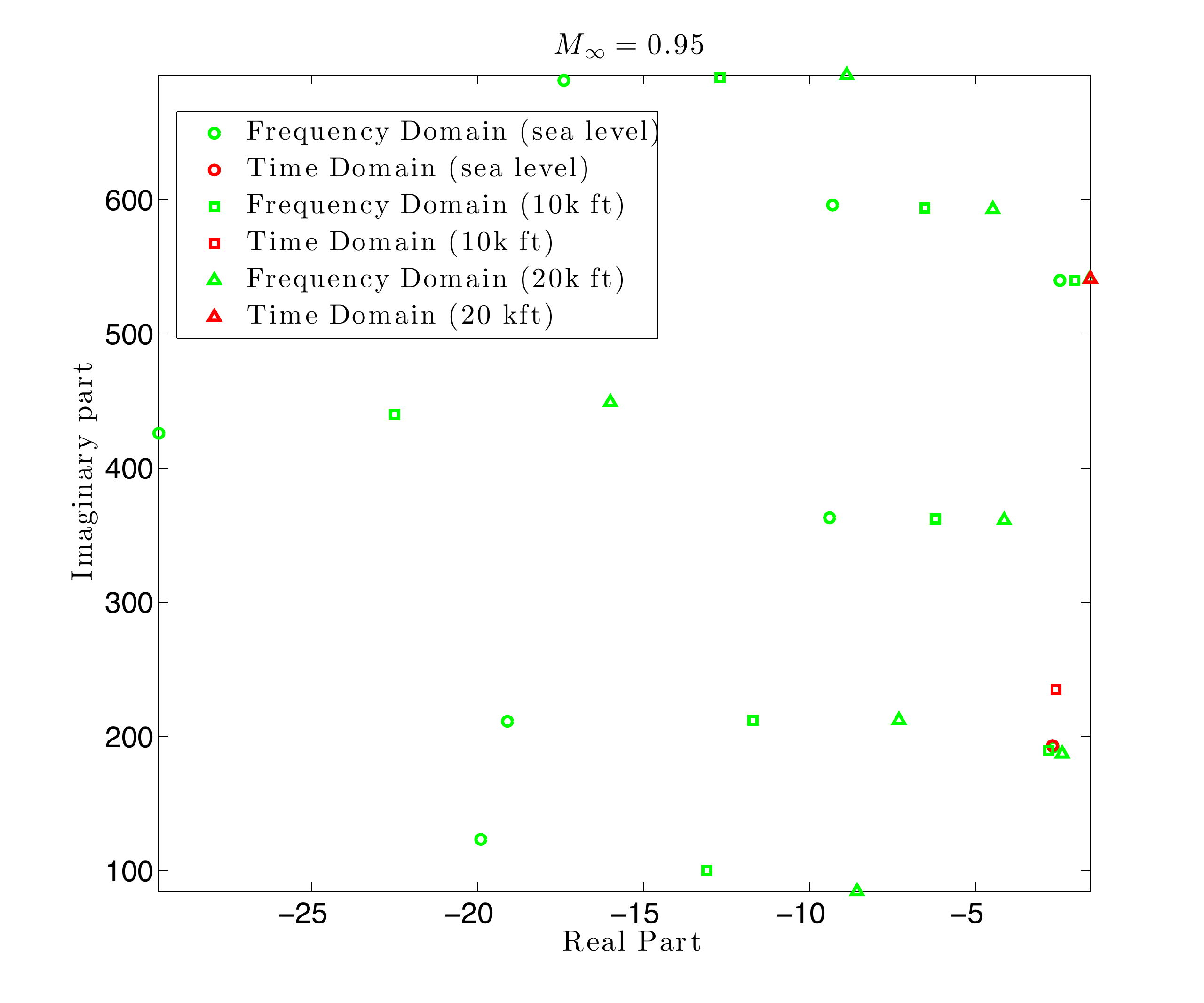}
}
\subfigure[$M_{\infty}=1.1$]{
\includegraphics[width=0.45\textwidth]{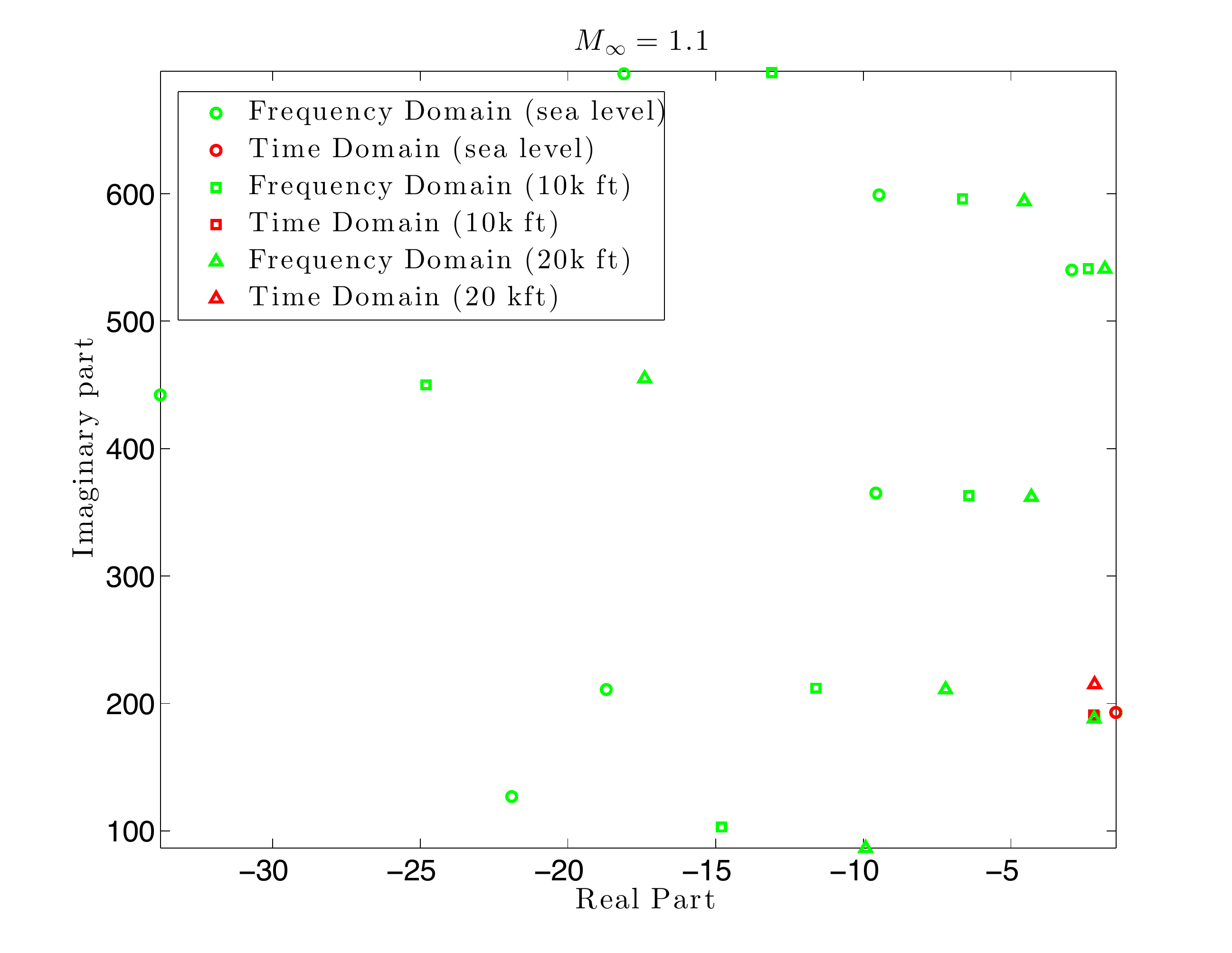}
}
\vspace{3mm}
\caption[]{
           Comparison of the aeroelastic eigenvalues predicted by the frequency and time domain methods for the ARW-2 wing
}\label{fig:arw2Evs}
\end{figure}

\begin{figure}[h!]
\vspace{-3mm}
\centering
\subfigure[$M_{\infty}=0.8$]{
\includegraphics[width=0.45\textwidth]{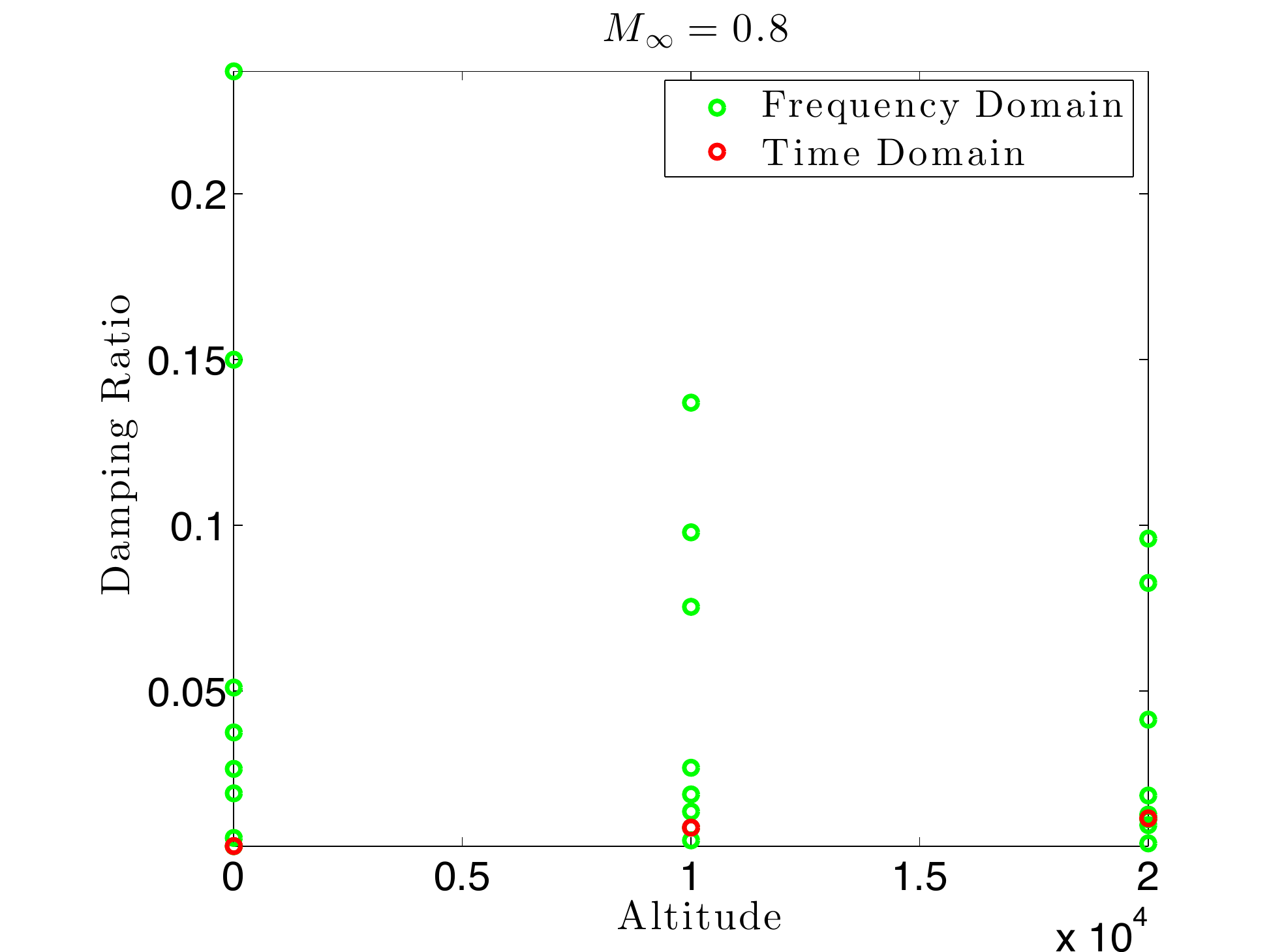}
}
\subfigure[$M_{\infty}=0.95$]{
\includegraphics[width=0.45\textwidth]{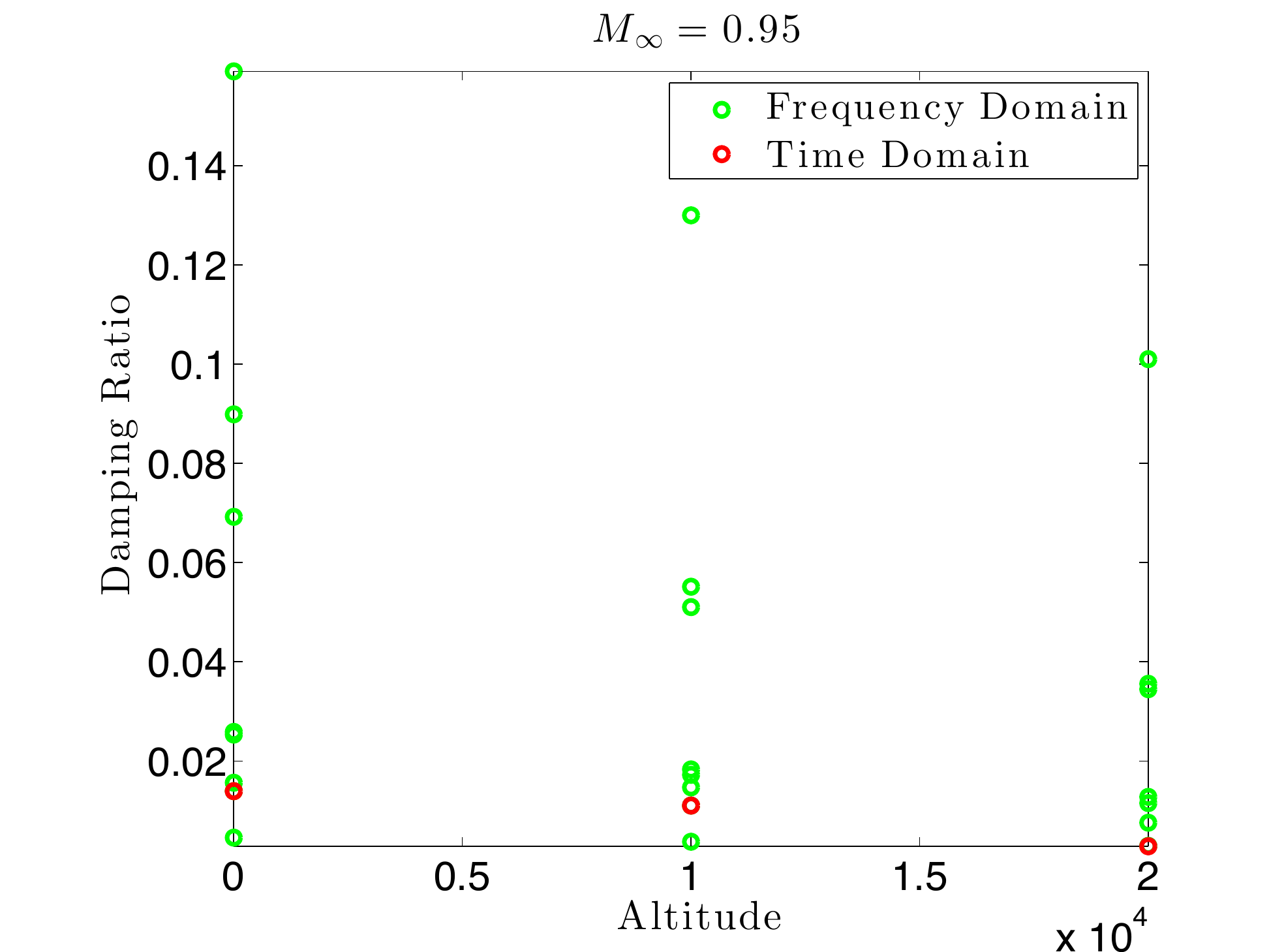}
}
\subfigure[$M_{\infty}=1.1$]{
\includegraphics[width=0.45\textwidth]{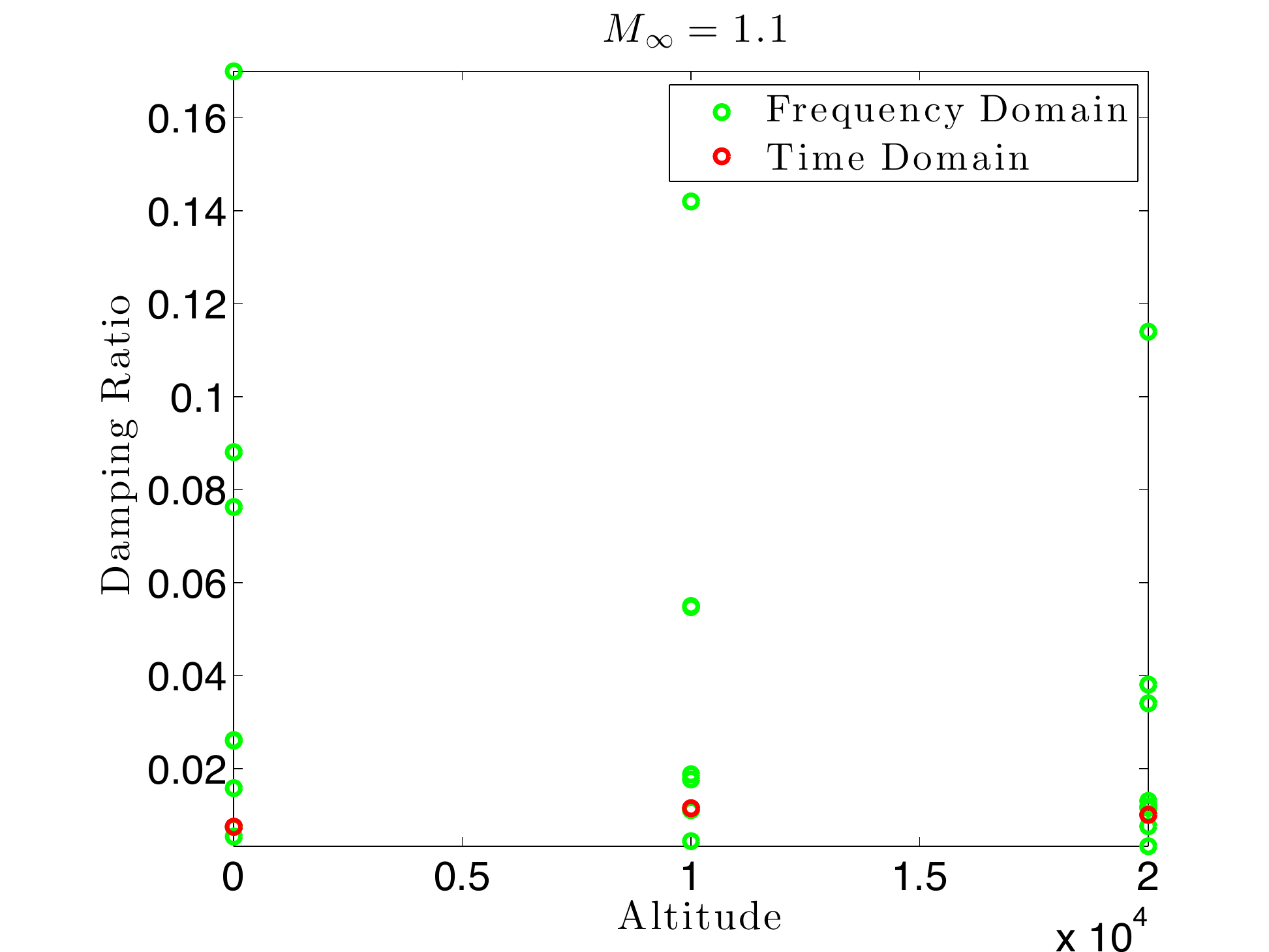}
}
\vspace{3mm}
\caption[]{
           Comparison of the smallest aeroelastic damping ratios predicted by the frequency and time domain methods for the ARW-2 wing
}\label{fig:arw2Damp} 
\end{figure}

Hence, for each case, eight sets of real and imaginary parts for the eigenvalues are computed.  The eigenvalues are reported in Figure~\ref{fig:arw2Evs}(a) for aeroelastic predictions at sea level, in Figure~\ref{fig:arw2Evs}(b) for $h=10,000$ ft and in Figure~\ref{fig:arw2Evs}(c) for $h=20,000$ ft, together with predictions in the time domain. One can observe that the time domain approach is again associated with large approximation errors of the aeroelastic eigenvalues, especially at $M_\infty=0.95$ and $M_\infty=1.1$. In turn, the minimal damping ratio is often mispredicted by this time-domain approach,as observed in Figure~\ref{fig:arw2Damp}.
The aeroelastic analysis for $M_\infty=0.95$ at sea level takes approximatively $6.8$ minutes on 32 processors for a tolerance $\epsilon=10^{-6}$. Computing the steady-state takes $1.2$ minute and the iterative eigenanalysis approach takes $5.6$ minutes.

\begin{figure}[h]
\centering
\includegraphics[width=0.6\textwidth]{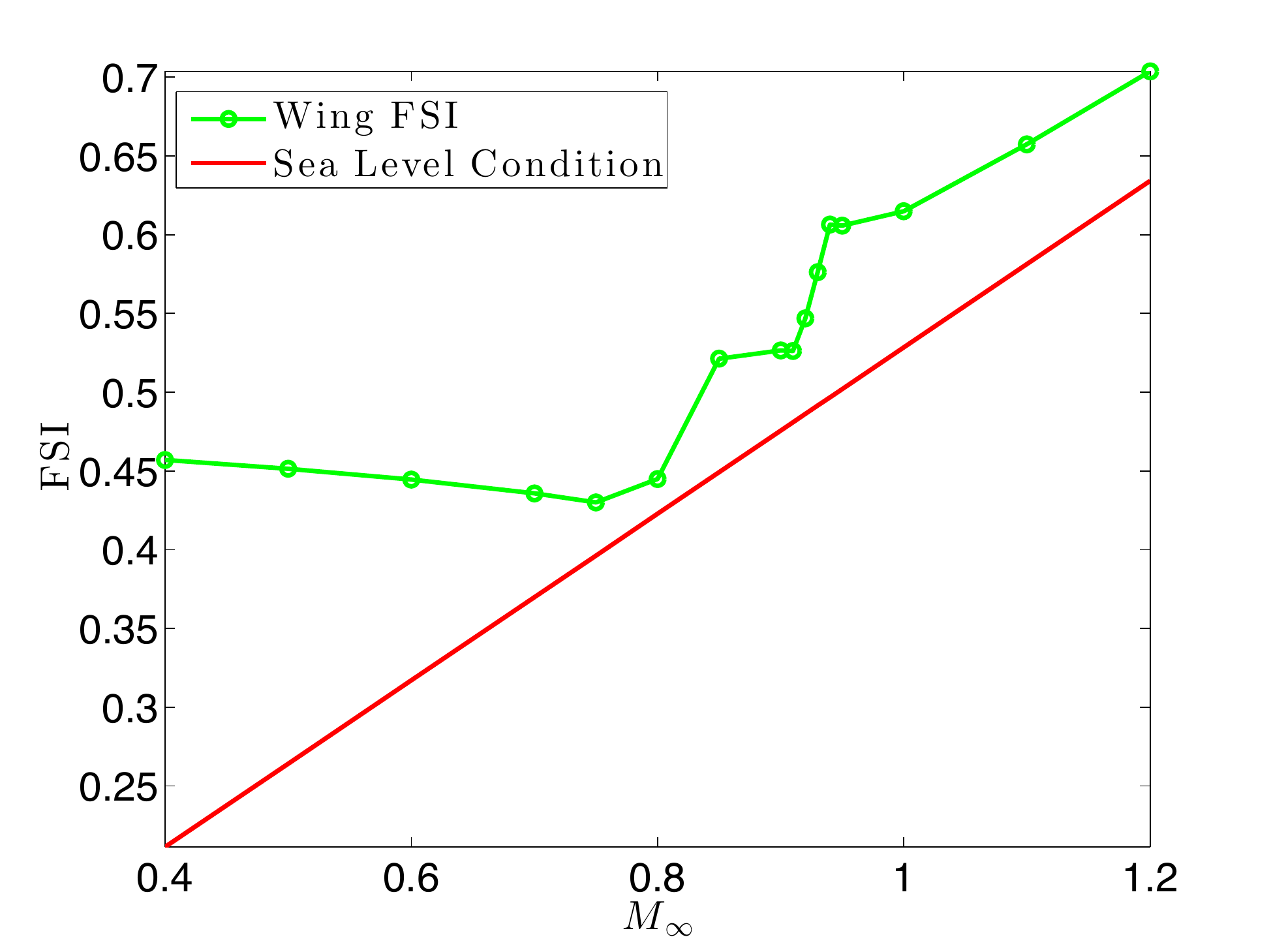}
\vspace{3mm}
\caption[]{
           Variations with the Mach number of the Flutter Speed Index and comparison with the sea level conditions for the ARW-2 wing
}\label{fig:FSIARW-2} 
\end{figure}

Finally, the flutter speed indices associated with the wing at sea level altitude are computed and reported in Figure~\ref{fig:FSIARW-2}. One can observe that flutter is never reached, even at sea level, confirming the proper flutter-free design of the ARW-2 wing.

\subsubsection{Design Optimization}
Next, the proposed aeroelastic analysis approach is included in a design optimization procedure in order to design an aircraft system under specific dynamic and static aeroelastic constraints. For that purpose, the ARW-2 wing is considered as a baseline design at $M_\infty=0.95$ and sea level altitude. In this study the first $m=6$ structural modes are considered.The following optimization problem is considered
\begin{gather}\label{eq:NLP}
\begin{split}
& \max_{\mubold\in\mathcal{D}} \frac{L(\mubold)}{D(\mubold)}\\
\text{s.t}~~~& W(\mubold) \leq W_{\max}\\
& \sigmabold(\mubold) \leq \sigma_{\max}\\
& \etabold(\mubold) \geq \eta_{\min},
\end{split}
\end{gather}
where $\mubold\in\mathbb{R}^{N_{\mubold}}$ denotes a vector of $N_{\mubold}$ design variables and $\mathcal{D}\subset \mathbb{R}^{N_{\mubold}}$ is the design space. $L(\mubold)$ and $D(\mubold)$ respectively denote the lift and drag associated with the static aeroelastic equilibrium for the design configuration $\mubold$. $W(\mubold)$ denotes the weight associated with the design $\mubold$. The vector $\sigmabold(\mubold)$ contains the nodal values of the von Mises stress in the structure at that configuration. Finally, $\etabold(\mubold)$ is the vector of $m$ modal damping ratios associated with the configuration $\mubold$ for the flight condition of interest. $W_{\max}$, $\sigma_{\max}$ and $\eta_{\min}$ provide respective bounds on the weight, stress and damping ratios. 

Tthere are $N_{\mubold}=6$ design variables. The first three are associated with the shape of the wing and respectively control the sweep, twist and dihedral angles. The last three variables control the thickness of three distinct subsets of spar and rib elements in the wing. The design space $\mathcal{D}$ is a box domain defined in Table~\ref{tab:boxDomain}. Note that all six design variables are non dimensional and scaled.

\begin{table}[htdp]
\begin{center}
\begin{tabular}{lclclclc|}
\hline
\hline
Design variable &  Nature & Lower bound & Upper bound \\
\hline
$s_1$& Sweep angle & $-0.1$ & $0.1$   \\
$s_2$& Twist angle & $-0.1$ & $0.1$  \\
$s_3$& Dihedral angle & $-0.1$ & $0.1$ \\
$s_4$& Element thickness & $-0.1$ & $0.1$ \\
$s_5$& Element thickness & $-0.1$ & $0.1$   \\
$s_6$&  Element thickness & $-0.1$ & $0.1$\\
\hline
\hline
\end{tabular}
\end{center}
\caption{Design domain for the ARW-2 wing}\label{tab:boxDomain}
\end{table}

The nonlinear program~(\ref{eq:NLP}) is solved with Matlab's fmincon active set method with convergence tolerances set to $10^{-6}$. The gradients are computed analytically except for the gradient of the damping ratio that is computed by finite differences. This requires a total of $N_{\mubold}+1=7$ calls to the iterative procedure introduced in this paper for each design considered.

\begin{figure}[h]
\centering
\includegraphics[width=0.9\textwidth]{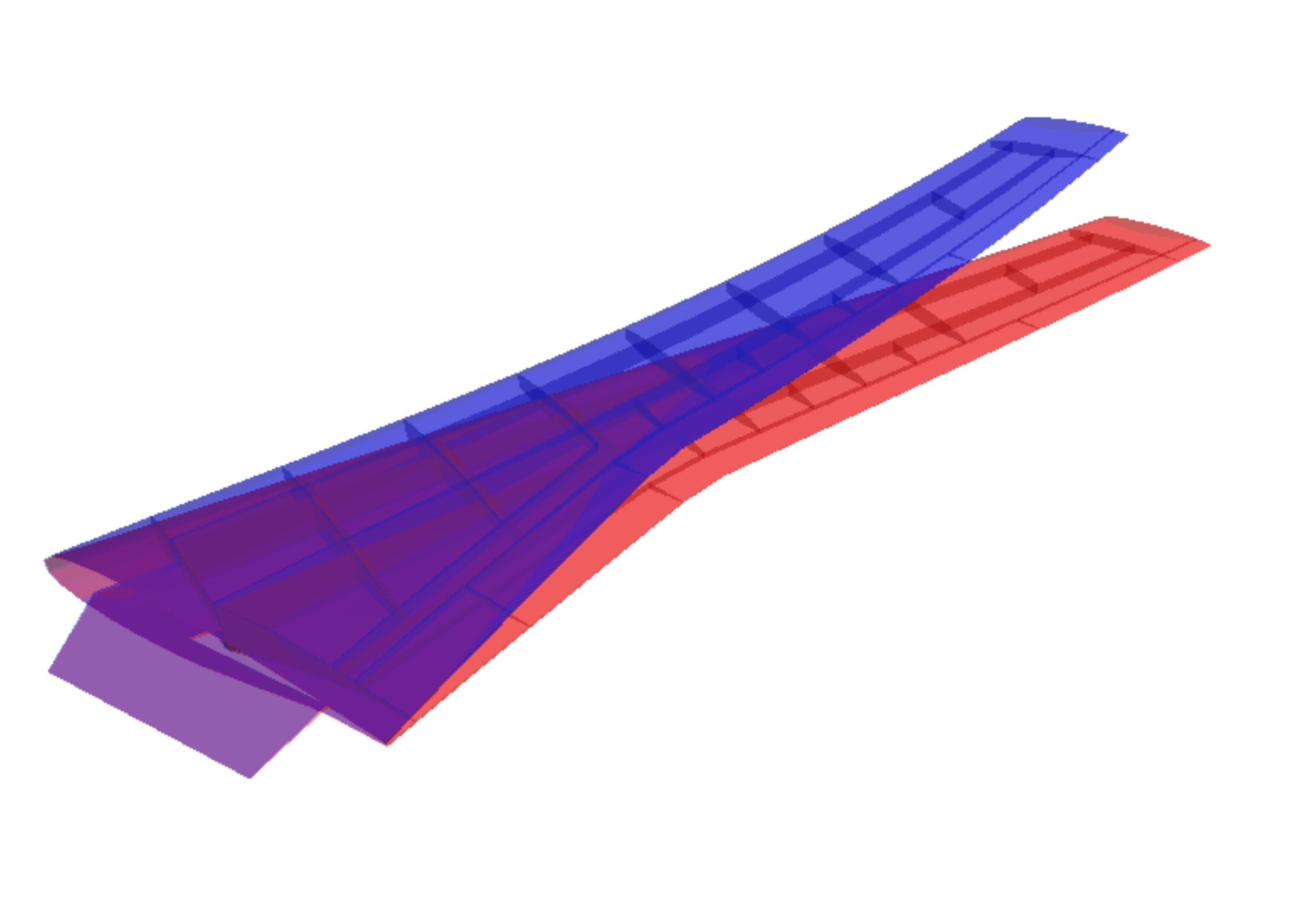}
\vspace{3mm}
\caption[]{
           Comparison of the initial design (in red) and the optimized design (in blue)
}\label{fig:designs}

\end{figure}

The initial design is the one associated with the original ARW-2 wing. Its physical properties are reported in Table~\ref{tab:designs}. The bounds in~(\ref{eq:NLP}) are set to $W_{\max}=400$ lbs, $\sigma_{\max}=2.5\times 10^{4}$ and $\eta_{\min}= 2\times10^{-4}$. As such, the dynamic aeroelastic constraint is violated for the initial design. The optimized design is found after 13 major iterations and reported in Table~\ref{tab:designs} and the two design configurations depicted in Figure~\ref{fig:designs}. Unlike the original design, the damping ratio constraint is satisfied and active for the optimized design. The lift/drag ratio has increased as well as the weight. The maximum von Mises stress has decreased.

\begin{table}[htdp]
\begin{center}
\begin{tabular}{lclclclc|}
\hline
\hline
Property &  Initial design & Optimized design \\
\hline
$s_1$&  $0$ & $-0.1$   \\
$s_2$& $0$ & $-0.1$  \\
$s_3$&  $0$ & $0.1$ \\
$s_4$&  $0$ & $0.1$ \\
$s_5$&  $0$ & $0.1$   \\
$s_6$&   $0$ & $0.011$\\
$\displaystyle{\frac{L(\mubold)}{D(\mubold)}}$&  $1.283$ & $1.355$ \\
$W(\mubold)$ (lbs)&  $349.9$ & $365.4$   \\
$\max(\sigmabold(\mubold))$ (psi)&   $1.54\times 10^{4}$ & $1.28\times 10^{4}$\\
$\min(\etabold(\mubold))$&  $1.82\times 10^{-4}$ & $2.00\times 10^{-4}$ \\
\hline
\hline
\end{tabular}
\end{center}
\caption{Properties of the initial and optimized designs}\label{tab:designs}
\end{table}

The  overall design procedure takes $8$ hours on 32 CPUs. About $1$ hour is associated with the computation of the multiple static constraint and its derivatives and $7$ hours to compute the dynamic constraints and their sensitivities. 

\begin{figure}[h]
\centering
\includegraphics[width=0.9\textwidth]{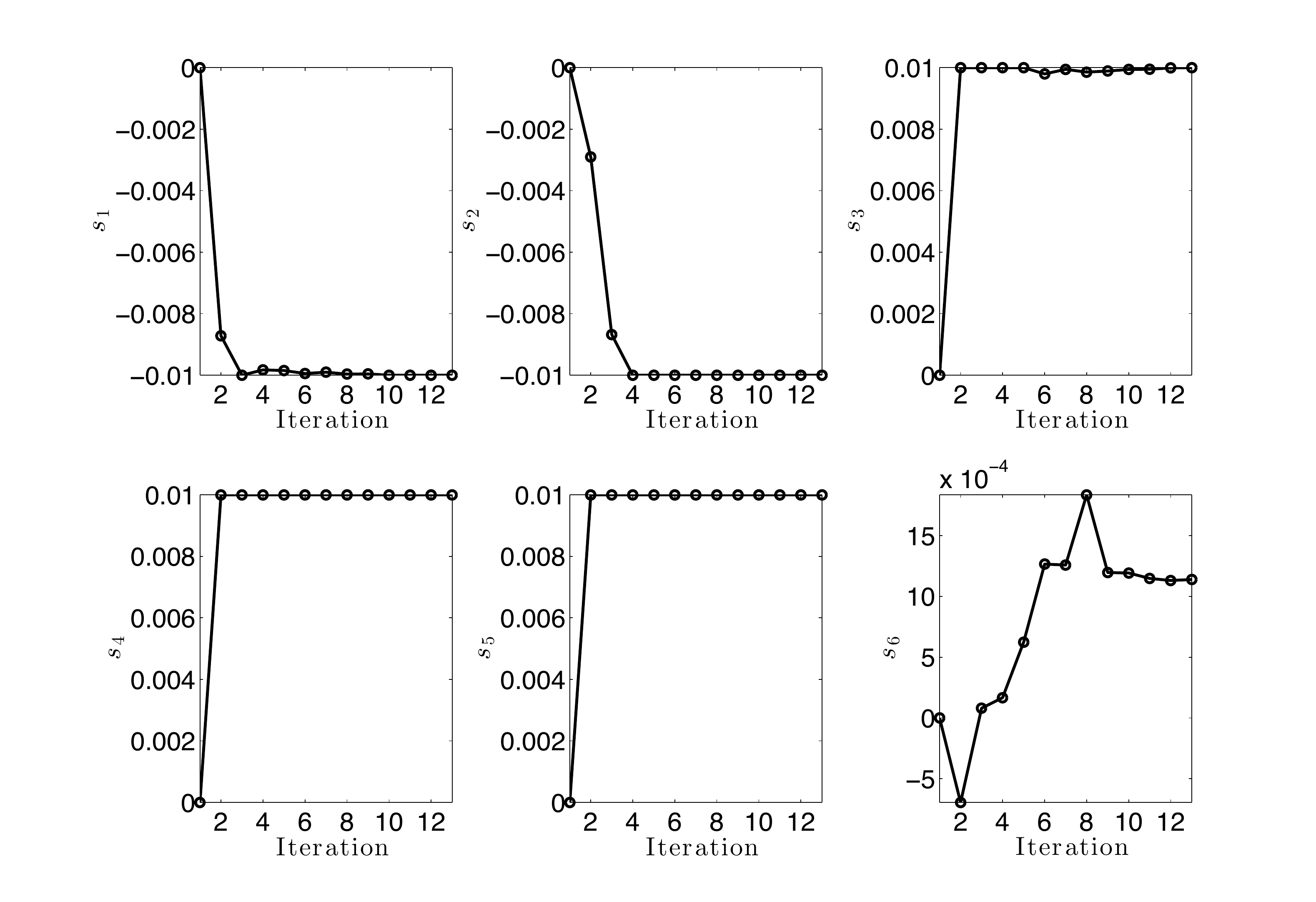}
\vspace{3mm}
\caption[]{
           Variations with the optimization iteration of the design variables
}\label{fig:iterations} 
\end{figure}

\begin{figure}[h]
\centering
\includegraphics[width=0.7\textwidth]{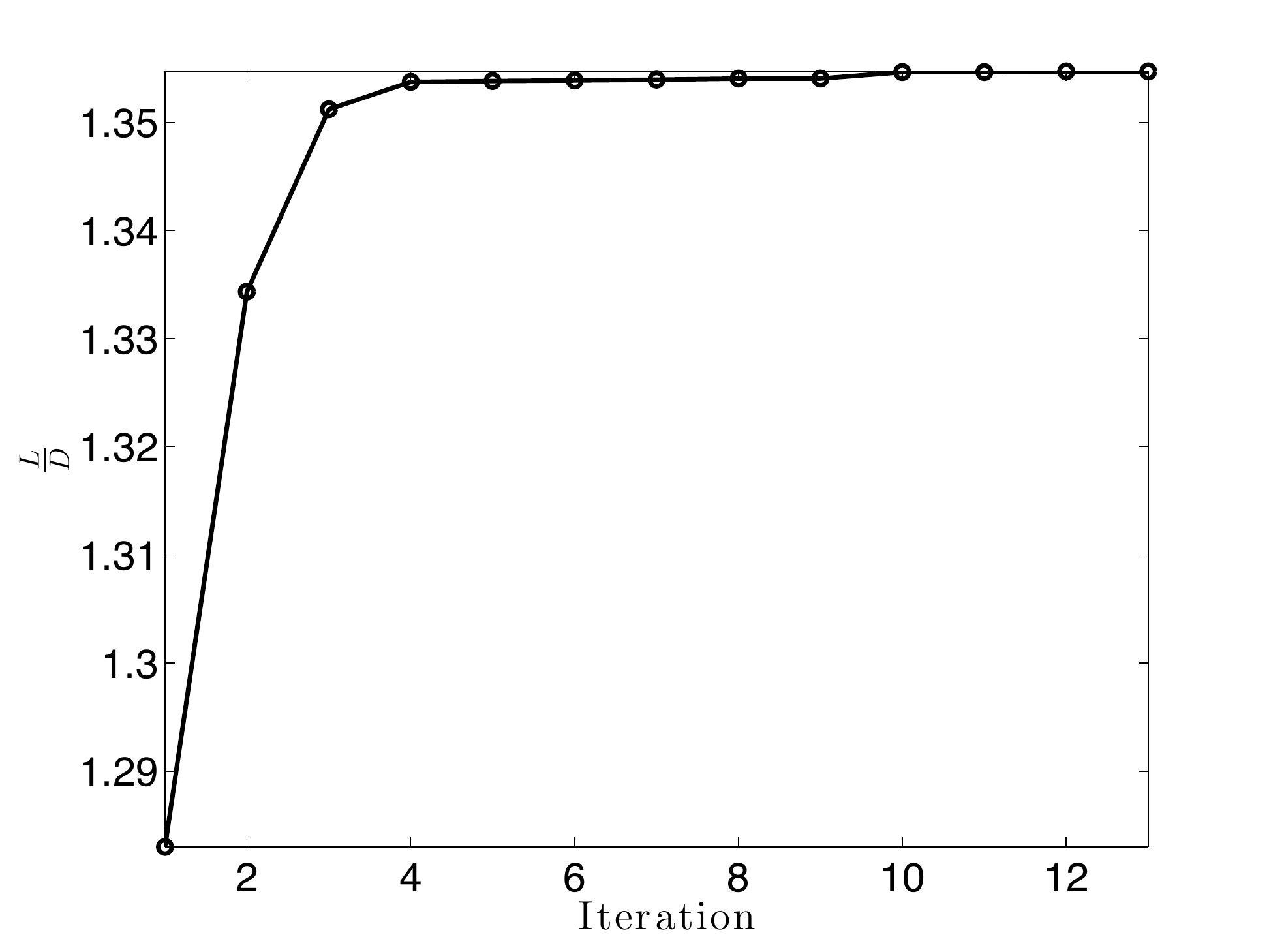}
\vspace{3mm}
\caption[]{
           Variations with the optimization iteration of the lift-to-drag ratio
}\label{fig:LDiterations} 
\end{figure}

The variation of the design over the optimization iterations is reported in Figure~\ref{fig:iterations} and the corresponding objective function in Figure~\ref{fig:LDiterations}.

\section{Conclusions}\label{sec:conclu}
This paper presents a practical and accurate approach for determining the aeroelastic characteristic of an aircraft configuration. The approach relies on a linearized CFD-based formulation together with fixed-point iteration procedure. The application of the proposed approach to the aeroelastic analysis and design optimization of two realistic wing configurations illustrates its  capability to enables fast and accurate aeroelastic predictions in the subsonic, transonic and supersonic domains.

\section{Bibliography}
\bibliography{manuscript}
\bibliographystyle{aiaa}
\end{document}